\documentclass{ptephy}

\usepackage{amsmath} 
\usepackage{amsthm} 
\usepackage{mathrsfs}
\usepackage{graphics} 
\usepackage{subfig} 

\renewcommand{\titlepagefootline}{\relax}
\renewcommand{\titlepageheadline}{\relax}

\begin{document}

\title{Gravitational radiation from an axion cloud around a black hole: Superradiant phase}

\author{\name{Hirotaka Yoshino}{1} and \name{Hideo Kodama}{1,2}}

\address{\affil{1}{Theory Center,
Institute of Particles and Nuclear Studies,
KEK, Tsukuba, Ibaraki, 305-0801, Japan}
\affil{2}{Department of Particle and Nuclear Physics,
Graduate University for Advanced Studies, Tsukuba 305-0801, Japan}
}

%
%
\begin{abstract}%
Motivated by possible existence of string axions with ultralight masses,
we study gravitational radiation from an axion cloud around a rotating
black hole (BH). The axion cloud extracts the rotation energy of the BH
by superradiant instability, while it loses energy
through the emission of gravitational waves (GWs). In this paper, GWs
are treated as perturbations on a fixed background spacetime 
to derive the energy emission rate.
We give an analytic approximate formula for the case where axion's Compton
wavelength is much larger than the BH radius,
and then, present numerical results without approximation.
The energy loss rate of the axion cloud
through the GW emission turns out to be smaller
than the energy gain rate of 
the axion cloud by superradiant instability 
until nonlinear self-interactions of axions
become important.
In particular, an axion bosenova must happen at the last stage
of superradiant instability.
\end{abstract}

\subjectindex{E31, E02, C15, B29}

\maketitle

%
%
\section{Introduction}

In a few years, the ground-based detectors,
Advanced LIGO, Advanced VIRGO, and KAGRA,
are expected to begin operation with sufficiently high sensitivity
to detect gravitational wave (GW) signals from
binary mergers of black holes (BHs) or neutron stars.
We will have a very exciting era of general relativity, and
many interesting sciences will be done.
One interesting possibility is
to find signals from unknown sources or effects
of unestablished hypothetical theories.

Do we have a possibility to detect signals from string theory
through GW detectors? Na\"ively it seems difficult because
the effect of string theory is expected to appear at very high energy
scales (e.g, the string scale).
However, the authors of Refs.~\cite{Arvanitaki:2009,Arvanitaki:2010}
answered {\it ``Maybe yes,''}
if there are string axions with ultralight masses.
Since string theories (or M-theory)
require our spacetime to have 10 or 11 dimensions,
the extra dimensions other than four have to be compactified.
Then, the extra dimensions have dynamical degrees of freedom
of changing their shape and size which are called moduli,
and they effectively behave as fields in our
four-dimensional spacetime. One of the moduli
will be the QCD axion. We can also expect the existence of
other pseudoscalar fields with ultralight masses, called string axions,
whose expected number is typically from 10 to 100. The allowed mass
range is from $10^{-10}$~eV to $10^{-33}$~eV and further below.
Depending on their masses,
they cause a variety of phenomena that could be observed in the context
of cosmophysics (=cosmology and astrophysics). This is called the axiverse scenario
(See also Ref.~\cite{Kodama:2011} for an overview).

Suppose one of the string axions has mass $\mu \sim 10^{-10}$~eV. The
Compton wavelength $1/\mu$ is comparable to the radius of a BH
with solar mass $M\sim M_{\odot}$
(Throughout this paper, we use the units $c=G=\hbar = 1$).
Then, around a rotating solar-mass BH,
there appears an unstable quasibound-state mode whose amplitude
grows exponentially through the extraction of the BH rotation
energy. This instability is called {\it superradiant instability}.
Due to this instability,
an axion cloud is formed around the BH
from quantum zero-point oscillations even if we start
from vacuum.
The growth rate
has been calculated by approximate methods \cite{Detweiler:1980,Zouros:1979}
and by numerical methods
\cite{Furuhashi:2004,Cardoso:2005,Dolan:2007}
by solving the frequency-domain eigenvalue problem.\footnote{The
growth rate was reproduced also in a time-domain
simulation by our code for a special case \cite{Yoshino:2012},
and more
systematic analyses of very long time evolutions were reported
by Dolan~\cite{Dolan:2012}.
See also \cite{Strafuss:2004} for an earlier work.}
The maximum growth rate has been found to be $M\omega_I\sim 10^{-7}$,
and the corresponding time scale to be
around $10^{7} M$, which is much longer than
$M$ (see Fig.~\ref{AC_growthrate}).
But this time scale is about $1$ minute for $M=M_{\odot}$.
Even for $M\omega_I \sim 10^{-12}$, the time scale is about $1$ day.
Therefore, for a wide range of parameters, the time scale of the
superradiant instability is much shorter than the observation period
of the ground-based GW detectors.

The axion cloud becomes denser and denser as the superradiant instability
proceeds, and two effects gradually become important. One
is nonlinear self-interaction, and the other is the GW emission.
Here, the nonlinear self-interaction comes from 
the potential of the axion field $\Phi$,
which is assumed to have the standard form 
$V = f_a^2\mu^2[1-\cos(\Phi/f_a)]$ 
in the present paper where $f_a$ is the axion decay constant.
In our previous paper \cite{Yoshino:2012},
we numerically simulated the time evolution of an axion field
obeying the sine-Gordon equation in the Kerr background spacetime.
The result is that the nonlinear self-interaction leads to
``axion bosenova,'' which shares some features with the
bosenova observed in experiments \cite{Donley:2001}
on Bose-Einstein condensates.
The bosenova is characterized by the
termination of superradiant instability and the gradual infall
of positive energy from the axion cloud into the BH. The bosenova
happens when the energy $E_a$ of the axion cloud  becomes
$E_a/M\approx 1600(f_a/M_p)^2$, where $M_p$ is the Planck mass.
If $f_a$ corresponds to the GUT scale, $f_a=10^{16}$~GeV,
the bosenova happens when axion cloud acquires 0.16\% energy
of the BH mass.

In the same paper \cite{Yoshino:2012},
we also gave an order estimate of the amplitude of GWs
emitted during the bosenova, and found a possibility to detect
signals from the bosenova if it happens around a BH near the solar system
(say, within 1 kpc). This motivates us to study GWs from the bosenova
in more detail. However, before doing that, we have to study GW
emission {\it before} the bosenova, i.e., in the superradiant phase.
The reason is as follows.
The axion cloud obtains energy by extraction of BH's rotation energy,
while it loses energy by emission of GWs.
The energy extraction rate $dE_a/dt$ is related to the superradiant growth
rate as
%
\begin{equation}
\frac{dE_a}{dt} = 2\omega_I E_a,
\label{superradiant-energy-extraction-rate}
\end{equation}
%
and therefore, it is proportional to $(E_a/M)$. On the other hand,
the energy loss rate by the GW emission (i.e., the radiation rate)
is $dE_{\rm GW}/dt \propto (E_a/M)^2$.
This is because the perturbation equation
indicates that the GW amplitude is proportional to $E_a$, and the
energy flux is proportional to the squared amplitude of GWs.
For this reason, $dE_a/dt$ is larger than $dE_{\rm GW}/dt$
when $E_a$ is sufficiently small. But there is a possibility
that $dE_{\rm GW}/dt$ may catch up with $dE_a/dt$ as $E_a$ is increased.
If this is the case, all of the extracted energy
from the BH is converted into GWs, and the growth of the axion cloud
stops. Then, bosenova does not happen
and the GW emission stays
at a level undetectable by GW experiments
in the near future. 
The GW emission rate in the linear growth phase also gives us 
the lower bound on the expected GW fluxes from potential sources.  
Therefore,
the study of GWs in the superradiant phase is important.

Approximate estimates of GW emissions from the BH-axion system were
previously given by Arvanitaki
{\it et. al} in Refs.~\cite{Arvanitaki:2009,Arvanitaki:2010}.
There, the GW radiation from
a huge gravitational atom in Newton potential was considered for
the case $M\mu\ll 1$.
The two GW emission processes were pointed out. One is
the radiation by level transitions of the axion cloud
as a gravitational atom. This process
occurs when the wavefunction of
the axion cloud is given by the superposition of two or
more levels of bound eigenstates, and GWs of the
frequencies identical to energy 
differences of the levels are radiated. The other
is the annihilation of two axions. This can happen even for an axion cloud
occupying just one bound-state level, and GWs with the frequency
$2\omega$ are emitted, where $\omega$ is the (real part of) the
axion bound-state frequency.
For both processes, they derived approximate formulas for the
GW radiation rates and concluded that
the radiation rates are sufficiently small to allow the growth 
of the axion cloud by the superradiant instability until
the occurrence of a bosenova.

In this paper, we consider the situation where
approximately all of the axion particles occupy one
bound-state level. Because the GW radiation
by the level transition is subdominant in such a situation, 
we focus on the two-axion
annihilation process.
We push forward the study of Ref.~\cite{Arvanitaki:2010} on this process
in two respects.
First, since the approximate analysis 
in Ref.~\cite{Arvanitaki:2010} was limited to the case where
the two axions to be annihilated 
are at the $2p$ level (i.e., $\ell=m=1$ and $n_r=0$), 
we extend to more general
cases of $\ell=m\ge 1$ and $n_r\ge 0$.
Next, we perform numerical calculations of the GW emission in
the Kerr background for general values of $M\mu$. This is certainly
necessary because the superradiant instability for the modes $\ell=m=2$
and $3$
becomes effective when $M\mu\sim 1$, where the approximation
$M\mu\ll 1$ breaks down.
Similarly to Ref.~\cite{Arvanitaki:2010}, the analysis is done within the
classical approximation.


This paper is organized as follows. In the next section, we describe
the general strategy and derive formulas that are used in the
subsequent two sections. In Sec.~3, we give an approximate
formula for the energy loss rate in the case $M\mu\ll 1$.
In this case, the flat spacetime can be adopted as the background spacetime, and we call this approximation the ``flat approximation''.
In Sec.~4, we explain the numerical method for the general value
of $M\mu$, where the Kerr spacetime is adopted as the background spacetime. 
Section~5 is devoted to a conclusion.
In Appendix~A, we describe technical details for computing
axion quasibound states around a Kerr BH numerically.
In Appendix~B, we sketch the derivation of the GW radiation rate
presented in Sec.~3.

%
%
\section{General strategy}

In this section, we briefly summarize the basic features 
of the superradiant growth of an axion cloud
around a rotating BH and the methods for
evaluating the GW radiation rate from this system.

\subsection{Superradiant bound states of an axion field}

We assume the BH at the center of the axion cloud to be
a rotating Kerr BH.
A Kerr BH is specified by the two parameters,
the mass $M$ and the angular momentum $J$.
The Kerr parameter $a$ is introduced by $J=Ma$.
In this paper, we often use the nondimensional rotation parameter
$a_*=a/M$. In the Boyer-Lindquist coordinates $(t, r, \theta, \phi)$,
the metric is given by
%
\begin{subequations}
\begin{multline}
ds^2 = -\frac{\Delta - a^2\sin^2\theta}{\Sigma}dt^2
-\frac{4Mar\sin^2\theta}{\Sigma}dtd\varphi\\
+\left[\frac{(r^2+a^2)^2-\Delta a^2\sin^2\theta}{\Sigma}\right]
\sin^2\theta d\varphi^2
+\frac{\Sigma}{\Delta}dr^2
+\Sigma d\theta^2,
\label{Metric-Kerr-ST}
\end{multline}
with 
\begin{equation}
\Sigma = r^2+a^2\cos^2\theta, \qquad \Delta = r^2+a^2-2Mr.
\end{equation}
\end{subequations}
%
The horizon is located at $r_+=M+\sqrt{M^2-a^2}$ and
it rotates with the angular velocity $\Omega_H = a/(r_+^2+a^2)$
as seen by observers at spatial infinity.

We consider a real scalar field $\Phi$ in the
Kerr spacetime. Here, $\Phi$ is treated as a test field
and its gravitational backreaction is neglected.
Because
we study the situation much before the bosenova,
the effect of the nonlinear self-interaction is less
important. For this reason, we ignore the nonlinear term
and assume $\Phi$ to satisfy
the massive Klein-Gordon equation
%
\begin{equation}
\nabla^2\Phi - \mu^2\Phi =0.
\label{massive-KG}
\end{equation}
%
In the Kerr spacetime, the separation of variables of
the field $\Phi$ is possible as
%
\begin{equation}
\Phi = \Re\hat{\Phi}, \qquad
\hat{\Phi} = 
e^{-i\omega t}R_{\ell m}^{~\omega \mu}(r)S_{\ell m}^{~\omega \mu}(\theta)e^{im\phi}.
\label{Phi-separation}
\end{equation}
%
Throughout the paper, $\hat{f}$ indicates the complexified
quantity of a real function $f$, i.e., $f=\Re\hat{f}$.
In the following, we denote $R=R_{\ell m}^{~\omega \mu}(r)$
and $S=S_{\ell m}^{~\omega \mu}(\theta)$ for simplicity. 
See Appendix~\ref{Appendix-A} for the 
equations that $S(\theta)$ and $R(r)$ satisfy.

The superradiant instability occurs only for quasibound states. 
Such a quasibound state is obtained by assuming the field to decay
at infinity and to be ingoing at the horizon.
The procedure is similar to the calculation of the bound states of
a hydrogen atom in quantum mechanics. In the case of the hydrogen atom,
one imposes the wave function to decay at infinity
and to be regular at the origin. Then, the energy levels are
discretized. In the same manner,
the frequency $\omega$ is discretized in the case of the BH-axion system.
Since there is an ingoing flux at the horizon,
the frequency $\omega$ becomes complex,
$\omega = \omega_R + i\omega_I$. The imaginary part gives the
exponential behavior, and the field decays for $\omega_I<0$ and
grows for $\omega_I>0$. In the case $\omega_I>0$, the bound state is unstable,
and this happens
when the superradiance condition
$\omega_R<m\Omega_H$ is satisfied. Physically, the instability happens
because the energy density in the neighborhood
of the BH horizon becomes negative for $\omega_R<m\Omega_H$, and
negative energy falls into the horizon.
Then, due to the energy conservation, the amplitude of
the outside field becomes larger.
This is the superradiant instability.

Approximate solutions for superradiant bound states were
constructed in the parameter region $M\mu\ll 1$ by the matched asymptotic
expansion (MAE) method \cite{Detweiler:1980}
and in the parameter region $M\mu\gg 1$ by the WKB method \cite{Zouros:1979}.
The solution in the MAE method is closely related to
our analysis in Sec.~3. In this method, two approximate solutions
are constructed in a near-horizon region and in a distant region,
and they are matched in the overlapping region where
both approximations are valid. The equation in the distant region
is identical to the Schr\"odinger equation for a hydrogen atom except that
the electric potential $-e^2/r$ is replaced by the Newton
potential $-M\mu/r$. Therefore,
the BH-axion system can be regarded as a huge gravitational atom,
which is characterized by the angular quantum numbers $(\ell, m)$
and the radial quantum number $n_r=0, 1, 2,...$.

%
\begin{figure}[tb]
\centering
\includegraphics[width=2.5in]{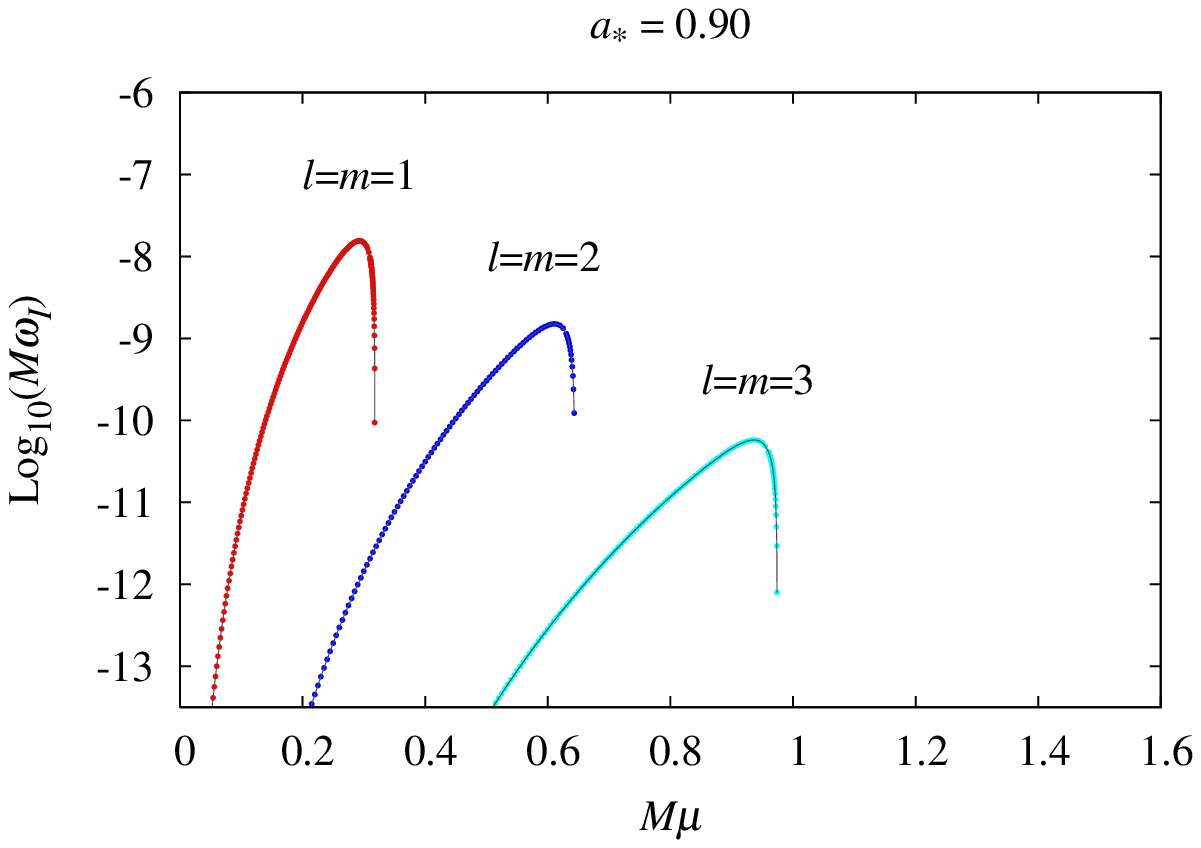}
\includegraphics[width=2.5in]{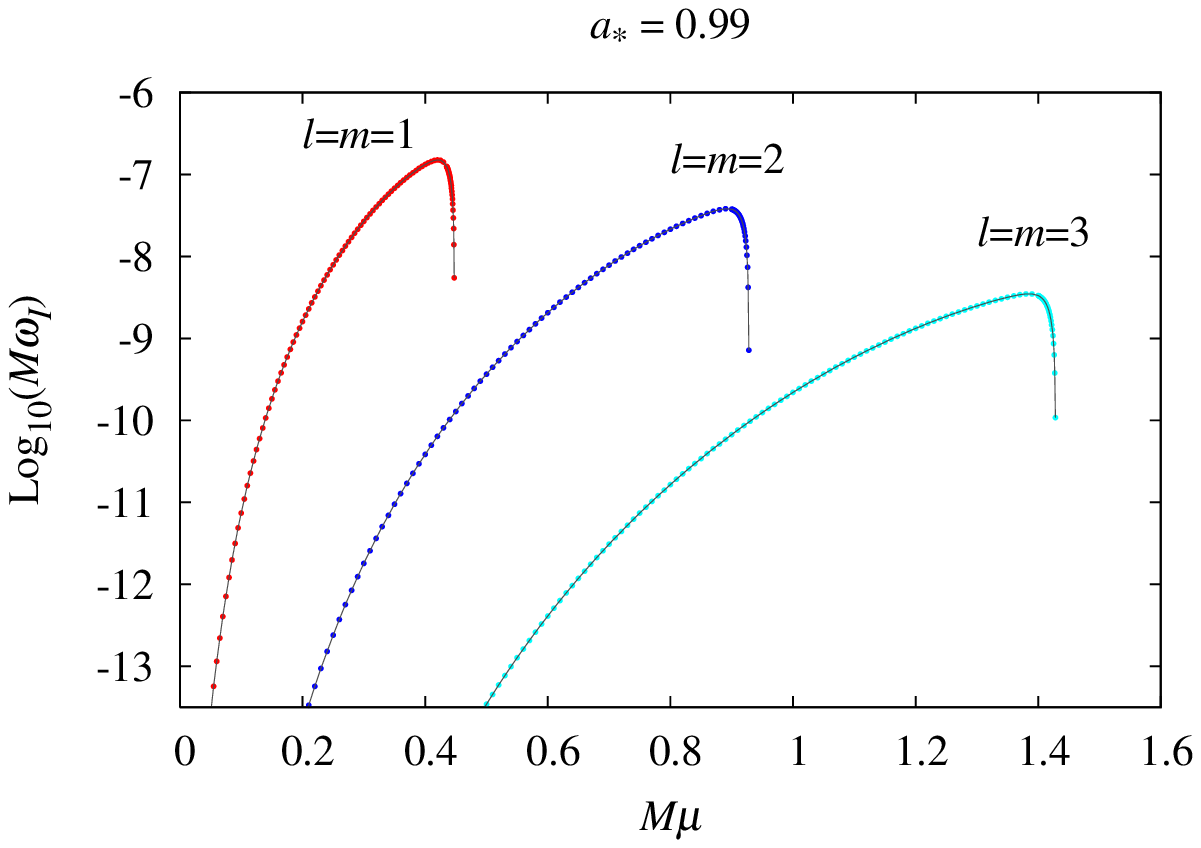}
\caption{Imaginary part $M\omega_I$ of bound state eigenfrequency as functions
of $M\mu$ for the BH rotation parameter $a_*=0.90$ (left) and $0.99$ (right). 
The results for the modes $\ell=m=1$, $2$, and $3$ are shown.}
\label{AC_growthrate}
\end{figure}
%

In the parameter range $M\mu\sim 1$, numerical calculations are required
to determine the bound-state levels.
Such analyses were done by several authors
\cite{Furuhashi:2004,Cardoso:2005,Dolan:2007}.
In Ref.~\cite{Dolan:2007}, a detailed analysis was reported
using the continued fraction method developed by Leaver \cite{Leaver:1985}.
We also reproduced the consistent result using our independent code.
See Appendix~\ref{Appendix-A} for technical details.
Figure~\ref{AC_growthrate} shows the imaginary part
$M\omega_I$ of the eigenfrequency as functions of $M\mu$ for
the BH rotation parameter $a_*=0.90$ and $0.99$. 
The results for the modes $\ell = m = 1, 2,$ and $3$
and $n_r=0$ are shown. The unstable range of $M\mu$ becomes
larger as $m$ is increased
because of the superradiance condition $\mu\approx\omega_R<m\Omega_H$.
The peak value of $M\omega_I$ becomes smaller as $\ell=m$ is increased
and as $a_*$ is decreased.

%
\begin{figure}[tb]
\centering
\includegraphics[width=2.6in]{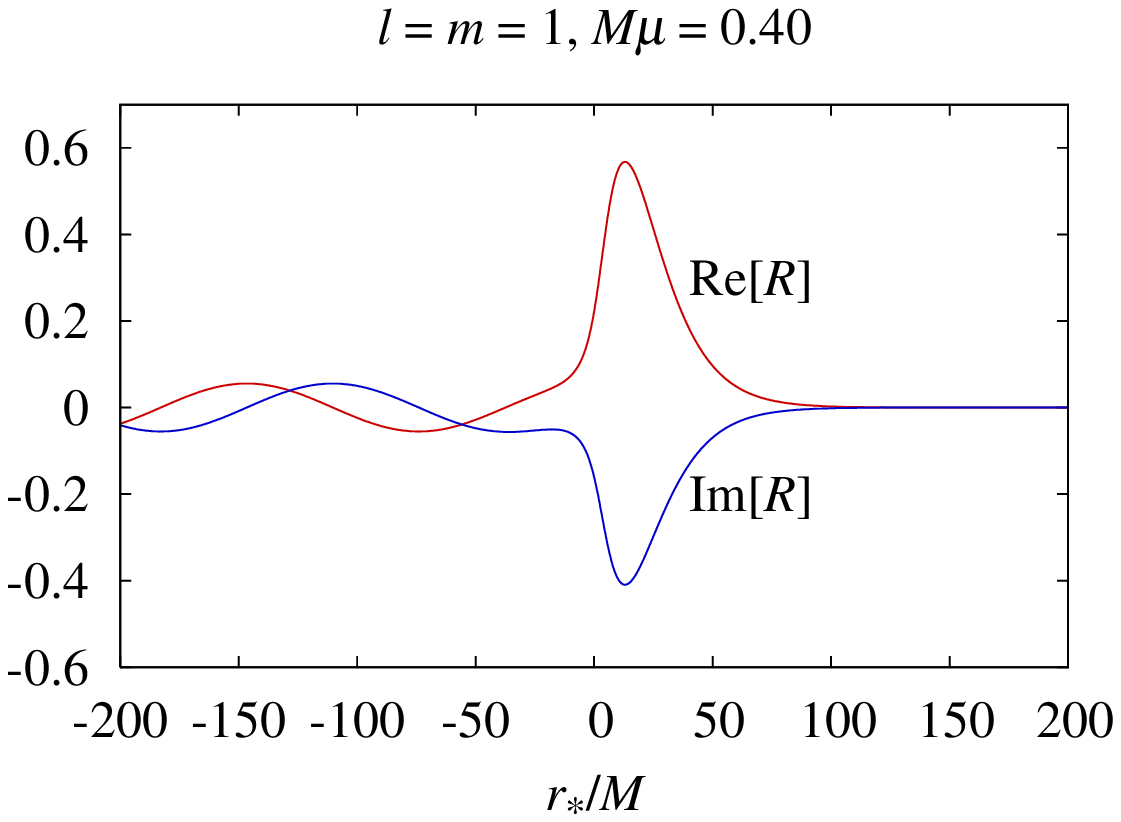}
\includegraphics[width=2.2in]{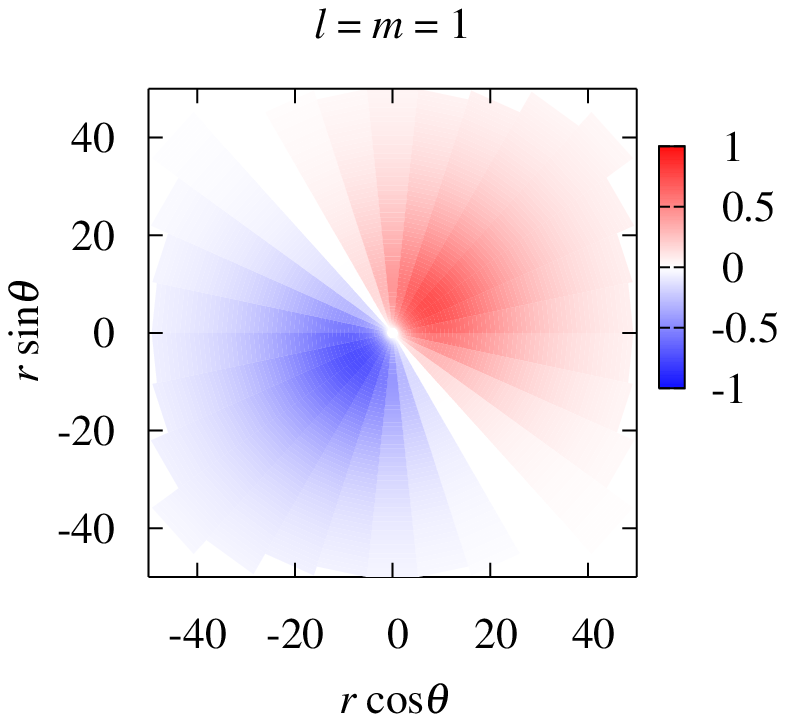}
\caption{Left panel: Radial function 
$R$ of the axion quasibound 
state as a function of the tortoise coordinate $r_*/M$
for $\ell = m = 1$,  $M\mu = 0.40$, and the BH
rotation parameter $a_*=0.99$. Right panel: The configuration
in the equatorial plane $\theta=\pi/2$. There are one maximum and 
one minimum.}
\label{AC_radial_L1M1}
\end{figure}
%

%
\begin{figure}[tb]
\centering
\includegraphics[width=2.6in]{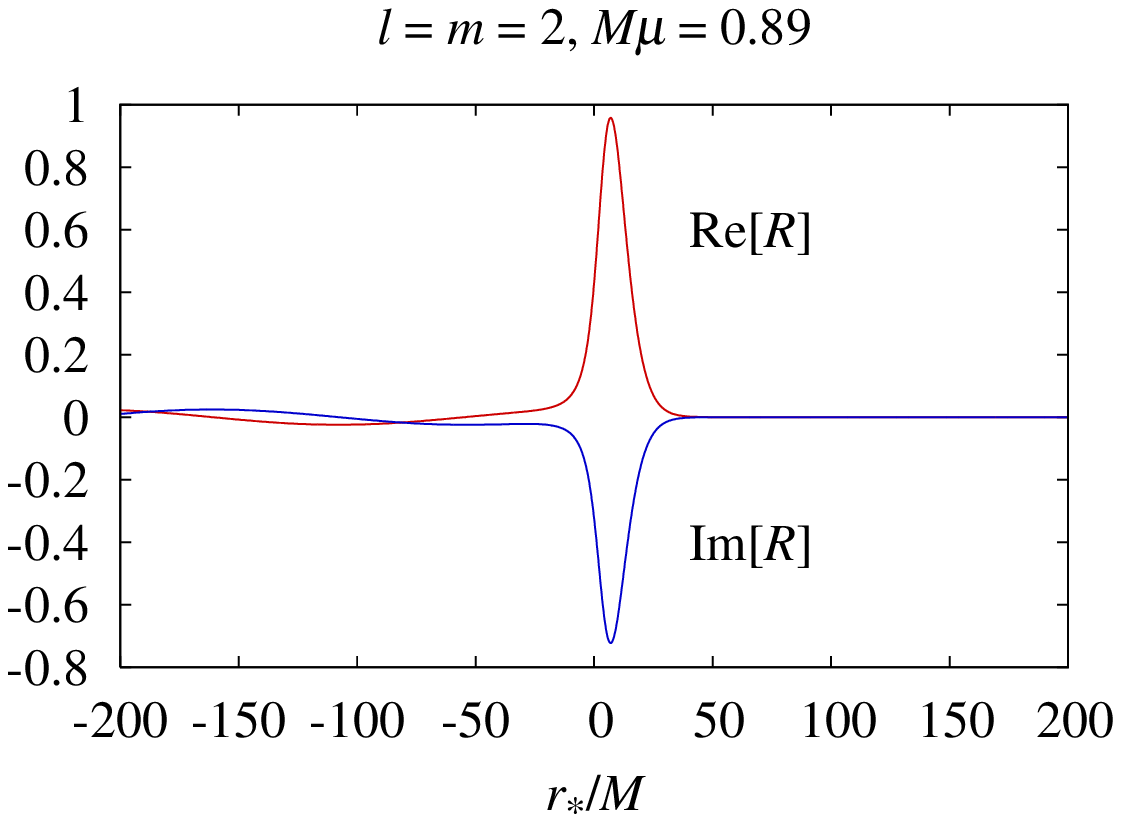}
\includegraphics[width=2.2in]{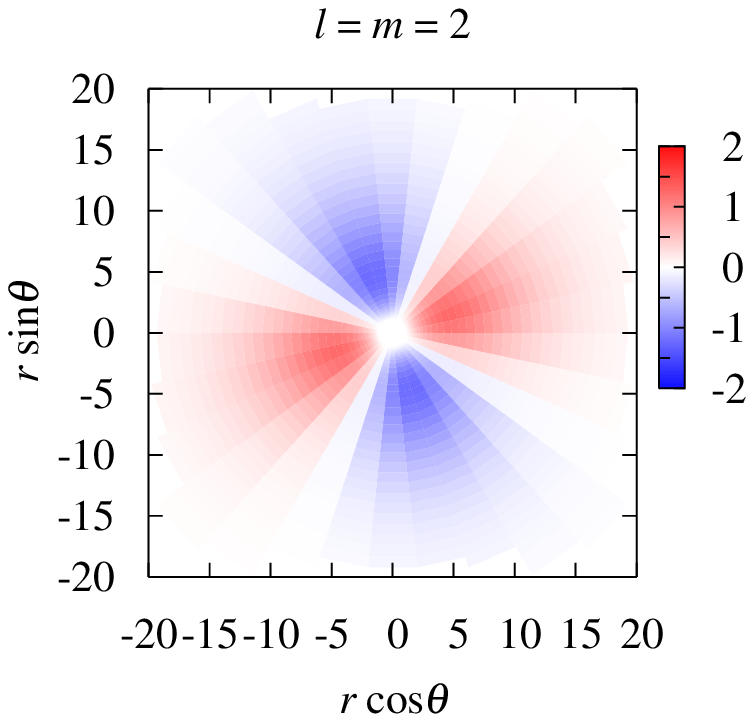}
\caption{The same as Fig.~\ref{AC_radial_L1M1} but for
$\ell = m = 2$ and  $M\mu = 0.89$. We can see
two maxima and two minima in the right panel.}
\label{AC_radial_L2M2}
\end{figure}
%

Figure~\ref{AC_radial_L1M1} shows an example of the radial profile of $R$
for the bound state solution 
with $\ell=m=1$ (the left panel) and the corresponding configuration
in the equatorial plane (the right panel). The system parameters
are adopted to be $M\mu=0.40$ and  $a_*=0.99$.
The function $R$
is shown as a function of the tortoise coordinate $r_*$ defined by
%
\begin{equation}
dr_* = \frac{r^2+a^2}{\Delta} dr.
\label{Def:tortoise_coordinate}
\end{equation}
%
Figure~\ref{AC_radial_L2M2} shows the same information 
for the mode with $\ell = m = 2$ and the system parameter $M\mu=0.89$.
In these two figures,
we chose the modes which are reduced to 
gravitational atoms with the radial quantum number $n_r=0$ for $M\mu\ll 1$.
For these modes, the field $\Phi$ has one maximum and one minimum 
for $\ell = m = 1$, and two maxima and two minima for $\ell = m = 2$ 
in the equatorial planes (i.e., right panels) 
outside of the ergoregion.
In the left panels, 
ingoing waves which behave as $R\sim e^{-i(\omega-m\Omega_H)r_*}$ can be
seen.
These waves carry negative energy density into the BH.
In Sec.~\ref{Sec:IV}, we use these numerical solutions
for mode functions of the quasibound states
as the source term of GWs.

\subsection{GWs from an axion cloud}

\label{Sec:II2}

The energy-momentum tensor of the axion field is
%
\begin{equation}
T_{\mu\nu} = T_{\mu\nu}(\Phi,\Phi),
\label{Energy-Momentum-Tensor-Original}
\end{equation}
%
where $T_{\mu\nu}(A,B)$ is defined as
%
\begin{equation}
T_{\mu\nu}(A,B) = \nabla_\mu A \nabla_\nu B
-\frac12 g_{\mu\nu}\left(\nabla_\rho A\nabla^\rho B+\mu^2 AB\right).
\label{Energy-Momentum-Tensor-AB}
\end{equation}
%
We study GWs generated by this energy-momentum tensor $T_{\mu\nu}$.
For this purpose, we consider the perturbation
%
\begin{equation}
\tilde{g}_{\mu\nu} = g_{\mu\nu} + h_{\mu\nu},
\end{equation}
%
where $\tilde{g}_{\mu\nu}$ is the spacetime metric,
$g_{\mu\nu}$ is the background Kerr metric, and $h_{\mu\nu}$
is a small perturbation. We introduce ${\psi}_{\mu\nu}$ by
%
\begin{equation}
{\psi}^{\mu}_{~\nu} = h^{\mu}_{~\nu}
-\frac12 \delta^\mu_{~\nu}h^{\rho}_{~\rho}.
\end{equation}
%
Then, in the de Donder gauge
%
\begin{equation}
\nabla^{\nu}{\psi}_{\mu\nu} = 0,
\end{equation}
%
the Einstein equations for ${\psi}_{\mu\nu}$
become
%
\begin{equation}
\Delta_L {\psi}_{\mu\nu} = 16\pi T_{\mu\nu},
\label{perturbative-eq1}
\end{equation}
%
where $\Delta_L$ is the Lichnerowicz operator that reduces in the vacuum case to
%
\begin{equation}
\Delta_L \varphi_{\mu\nu} = -\nabla^2\varphi_{\mu\nu}
-2R_{\mu ~ \nu}^{~\rho~\sigma}\varphi_{\rho\sigma}.
\label{perturbative-eq2}
\end{equation}
%
Equation~\eqref{perturbative-eq1} can be rewritten
as the equation for $h_{\mu\nu}$,
%
\begin{equation}
\Delta_L {h}_{\mu\nu} 
= 
16\pi \left(T_{\mu\nu}-\frac{1}{2} T^{\rho}_{~\rho}g_{\mu\nu}\right).
\label{perturbative-eq3}
\end{equation}
%

Here, we introduce one approximation.
The field $\Phi$ is proportional to $e^{\omega_I t}$
by the superradiant instability, and
the energy-momentum tensor gradually grows larger.
However, the time scale of the instability is very long, $\gtrsim 10^7M$,
and its effect on the GW flux must be small.
For this reason, we ignore the imaginary part $\omega_I$ and calculate
GWs from an oscillating source without
exponential growth. This also means that we ignore the
ingoing flux of the field $\Phi$ to the BH horizon.

In addition to this, there is another subtlety. 
Originally, the superradiant unstable mode for $\Phi$ is obtained 
as a complex function $\hat \Phi$ proportional to $e^{-i\omega t + im\phi}$.  
Then, the energy-momentum tensor should be estimated 
by using its real part $\Phi=\Re{\hat \Phi}$ as
%
\begin{equation}
T_{\mu\nu}(\Phi,\Phi)
=\frac14 
\left[ 
T_{\mu\nu}(\hat\Phi,\hat\Phi)
+ T_{\mu\nu}(\hat\Phi^*,\hat\Phi^*)
+2T_{\mu\nu}(\hat\Phi,\hat\Phi^*)
\right],
\label{Energy-Momentum-Real-Complex}
\end{equation}
%
where $z^*$ denotes the complex conjugate of $z$.
Clearly, the left-hand side of Eq.~\eqref{Energy-Momentum-Real-Complex} 
is not the real part of $T(\hat\Phi,\hat\Phi)/2$ 
when $\hat{\Phi}$ are sum of several oscillating modes 
proportional to $e^{-i(2\omega)t + i(2m)\phi}$ because of
the presence of the term $T(\hat\Phi,\hat\Phi^*)$.
In fact, this term can be interpreted as the source
for the GW emission by the level transition \cite{Arvanitaki:2010}. 
However, for a monochromatic eigenmode of $\Phi$, 
the stationary terms corresponding to $T(\hat\Phi,\hat\Phi^*)$ 
generate only stationary perturbations $h_{\mu\nu}$, 
and we are not interested in such perturbations. 
Hence, we neglect this contribution and focus on 
GWs generated by the oscillating terms of 
Eq.~\eqref{Energy-Momentum-Real-Complex}. 
For this purpose,
it is convenient to introduce
the complexified energy momentum tensor $\hat{T}$ defined by
%
\begin{equation}
T_{\mu\nu}(\Phi,\Phi) = \Re(\hat{T}_{\mu\nu}) 
\quad \textrm{with} \quad 
\hat{T}_{\mu\nu}=(1/2)T_{\mu\nu}(\hat\Phi,\hat\Phi).
\end{equation}
%
Since $\hat{T}$ is proportional to $e^{-2i\omega t+2im\phi}$, 
the frequency 
$\tilde{\omega}$ and the azimuthal quantum number $\tilde{m}$ 
of GWs satisfy
%
\begin{equation}
\tilde{\omega} = 2\omega, \qquad \tilde{m} = 2m.
\label{general-frequency-azimuthal}
\end{equation}
%
Hereafter, tilder indicates a quantity for GWs.

Once a solution for the mode function is given,
the GW radiation rate toward null infinity can be calculated by
%
\begin{equation}
\frac{dE^{\rm (out)}_{\rm GW}}{dt} = \frac{1}{32\pi}
\int_{r=r_{\rm out}}
g^{\mu\rho}g^{\nu\sigma}
\dot{h}_{\mu\nu}^{\mathrm{TT}}
\dot{h}^{\mathrm{TT}}_{\rho\sigma} dS,
\label{Eq:radiation-efficiency}
\end{equation}
%
where $\dot{h}_{\mu\nu}^{\mathrm{TT}}$ denotes the time derivative
of a metric perturbation
in the transverse-traceless (TT) gauge, 
and $dS$ is the area element of the 2-surface in a distant region
specified by $t$ and $r=r_{\rm out}$. 
Here, the conditions for the TT gauge are
%
\begin{equation}
\nabla^\mu h^{\textrm{TT}}_{\mu\nu}=0
\quad
\textrm{and}
\quad
g^{\mu\nu} h^{\textrm{TT}}_{\mu\nu}=0,
\end{equation} 
%
and the existence of the TT gauge has been shown for
a vacuum perturbation of a vacuum spacetime (e.g., p.~186 of Ref.~\cite{Wald}).
On the other hand,
the formula for the energy flux $dE_{\rm GW}^{\rm (bh)}/dt$
absorbed by the horizon of a Kerr BH  
is derived in Ref.~\cite{Teukolsky:1974}
in terms of the Newman-Penrose quantity
$\psi_0$ or $\psi_4$. A consistent formula is found 
using the Isaacson effective energy-momentum tensor for high-frequency GWs
in Ref.~\cite{Chrzanowski:1975}. 
In this paper, we do not explicitly calculate  
the GW energy flux across the horizon with the following reason:  
In Refs.~\cite{Teukolsky:1974,Chrzanowski:1975}, 
$dE_{\rm GW}^{\rm (bh)}/dt$ has been shown to be proportional to 
$\tilde{\omega}-\tilde{m}\Omega_H$. As stated in
Eqs.~\eqref{general-frequency-azimuthal}, the frequency and the azimuthal
quantum number of GWs satisfy $\tilde{\omega}=2\omega$ and $\tilde{m}=2m$.
Then, if the axion cloud satisfies the superradiance condition
$\omega<m\Omega_H$, emitted GWs also satisfy the superradiance condition
$\tilde{\omega}<\tilde{m}\Omega_H$. Therefore, the GW energy flux
to the horizon is negative, ${dE_{\rm GW}^{\rm (bh)}}/{dt}<0$,
and hence, the total radiation rate is
%
\begin{equation}
\frac{dE_{\rm GW}}{dt}
=
\frac{dE_{\rm GW}^{\rm (out)}}{dt}+\frac{dE_{\rm GW}^{\rm (bh)}}{dt}
<\frac{dE_{\rm GW}^{\rm (out)}}{dt}.
\end{equation}
%
This means that ${dE_{\rm GW}^{\rm (out)}}/{dt}$ gives the upper
bound of the total energy loss rate ${dE_{\rm GW}}/{dt}$ 
of the axion cloud due to the
GW emission, and if ${dE_{\rm GW}^{\rm (out)}}/{dt}$ is smaller
than the energy extraction rate $dE_a/dt$ of the axion cloud, 
we can safely declare that
the GW emission does not stop the
growth of the axion cloud by consuming all the energy extracted from the BH by
the superradiant instability.

In order to estimate the GW emission rate 
toward null infinity with the aid of the 
formula~\eqref{Eq:radiation-efficiency}, 
in principle, we have to solve Eq.~\eqref{perturbative-eq1} first 
and then find a gauge transformation that puts the solution into the TT gauge. 
However, this is a rather intricate task. 
Therefore, we adopt the following different method.

First, we pay attention to the fact that the solution $h_{\mu\nu}$ 
of Eq.~\eqref{perturbative-eq3} satisfies the vacuum Einstein equations 
outside the finite source region. 
Therefore, 
it is related to the homogeneous out-mode solutions\footnote{
The homogeneous GWs are classified
into four independent modes: in-, out-, up-, and down-modes
depending on the boundary conditions at future and past null infinity
$\mathscr{I}^+$ and $\mathscr{I}^-$ 
and the future and past horizons $H^+$ and $H^-$
(see, e.g., p.~93 of Ref.~\cite{Frolov:1998}).} $\hat{u}^{(\tilde{j})}_{\mu\nu}$
at future null infinity $\mathscr{I}^+$ 
as
%
\begin{equation}
\hat{h}_{\mu\nu} = 
\sum_{\tilde{j}}C_{\tilde{j}}\hat{u}^{(\tilde{j})}_{\mu\nu} 
\quad \textrm{at} \quad 
\mathscr{I}^+.
\label{expand-hmunu-scriplus}
\end{equation}
%
Here, we introduce a complex metric perturbation
$\hat{h}_{\mu\nu}\propto 
e^{-i\tilde{\omega}t+i\tilde{m}\phi}$
with $h_{\mu\nu} =\Re[\hat{h}_{\mu\nu}]$ as before, and 
$\tilde{j}$ indicates the collection of indexes to specify
each mode, i.e. 
the angular quantum numbers $\tilde{\ell}$ and $\tilde{m}$ 
and the ``polarization state'' $P=\pm 1$
(in the definition of Chrzanowski~\cite{Chrzanowski:1975}). 
The out-mode $\hat{u}^{(\tilde{j})}_{\mu\nu}$ is 
a solution to the homogeneous equation
%
\begin{equation}
\Delta_L \hat{u}^{(\tilde{j})}_{\mu\nu} = 0
\end{equation}
%
satisfying the boundary condition of vanishing flux 
across the future horizon $H^+$ (i.e., the absence 
of ingoing waves at $r_*\to -\infty$). Since we consider monochromatic waves, 
the out-mode solutions $\hat{u}^{(\tilde{j})}_{\mu\nu}$ are also
in proportion to $e^{-i\tilde{\omega}t+i\tilde{m}\phi}$. 
By a gauge transformation, the perturbation can be
expressed in the TT gauge, 
in which $\hat{h}_{\mu\nu}^{\rm TT}$ can be expanded 
in terms of $\hat{u}^{(\tilde{j}) \mathrm{TT}}_{\mu\nu}$ 
with the same expansion coefficients 
$C_{\tilde{j}}$. 
By inserting this expansion in the TT gauge 
into \eqref{Eq:radiation-efficiency}, we obtain
%
\begin{equation}
\left\langle \frac{dE^{\rm (out)}_{\rm GW}}{dt} \right \rangle
  = \frac{\tilde{\omega}^2}{64\pi}\sum_{\tilde{j}}\left|C_{\tilde{j}}\right|^2
  J_{\tilde{j}}
 \label{PGW:general}
\end{equation}
%
with
%
\begin{equation}
J_{\tilde{j}} 
= 
\int g^{\mu\rho}g^{\nu\sigma}\hat{u}_{\mu\nu}^{(\tilde{j})\mathrm{TT}}
\hat{u}^{*(\tilde{j})\mathrm{TT}}_{\rho\sigma}dS,
\label{def:J_tildej}
\end{equation}
%
where $\left\langle X \right \rangle$ implies 
the time average over a time sufficiently larger than $1/\tilde\omega$.

Next, we determine $C_{\tilde{j}}$. 
First, we apply Green's theorem to the identity
%
\begin{equation}
\nabla_\rho\left(\hat{u}^{*(\tilde{j})\mathrm{TT}}_{\mu\nu}\nabla^\rho \hat{h}^{\mu\nu}
-\hat{h}^{\mu\nu}\nabla^\rho \hat{u}^{*(\tilde{j})\mathrm{TT}}_{\mu\nu}\right)
=-16\pi \hat{u}^{*(\tilde{j})\mathrm{TT}}_{\mu\nu}\hat{T}^{\mu\nu}
\end{equation}
%
to obtain
%
\begin{equation}
\int_{\partial D}\left(\hat{u}^{*(\tilde{j})\mathrm{TT}}_{\mu\nu}\nabla^\rho \hat{h}^{\mu\nu}
-\hat{h}^{\mu\nu}\nabla^\rho \hat{u}^{*(\tilde{j})\mathrm{TT}}_{\mu\nu}\right)
n_\rho d\Sigma
=-16\pi \int_D\hat{u}^{*(\tilde{j})\mathrm{TT}}_{\mu\nu}\hat{T}^{\mu\nu}\sqrt{-g}d^4x.
\label{Eq:integral-relation1}
\end{equation}
%
This relation holds for an arbitrary domain $D$. 
Let us adopt as $D$ the domain surrounded by 
$t=t_0$ and $t=t_0+\Delta t$, $r=r_{\rm out}$, and $r=r_{\rm in}$. 
Here, $r=r_{\rm out}$ is located at a sufficiently far position
from the axion cloud, and $r=r_{\rm in}$ is located close to the horizon.

In the left-hand side of the relation \eqref{Eq:integral-relation1}, 
the surface integrals at $t=t_0$ and $t=t_0+\Delta t$ remain finite and 
we can safely neglect their contribution if we take a sufficiently large 
$\Delta t$. 
Then, the both sides of this relation are proportional to $\Delta t$. 
Besides, because of the boundary condition for 
${u}^{*(\tilde{j})\mathrm{TT}}_{\mu\nu}$
imposed at the future horizon, the surface integral at
$r=r_{\rm in}$ vanishes. 
Hence, we have
%
\begin{equation}
-16\pi\langle u^{(\tilde{j})\mathrm{TT}}, T\rangle
=\int_{r=r_{\rm out}}
\left(\hat{u}^{*(\tilde{j})\mathrm{TT}}_{\mu\nu}\partial_{r} h^{\mu\nu}
-h^{\mu\nu}\partial_{r} {u}^{*(\tilde{j})\mathrm{TT}}_{\mu\nu}\right)
 dS
\label{Eq:integral-relation2}
\end{equation}
%
where the ``inner product'' of ${u}_{\mu\nu}$ and ${T}_{\mu\nu}$
is defined as
%
\begin{equation}
\langle {u}, {T}\rangle
=\frac{1}{\Delta t}\int_{t_0}^{t_0+\Delta t}dt
\int \hat{u}_{\mu\nu}\hat{T}^{\mu\nu}\sqrt{-g}drd\theta d\phi.
\label{Eq:definition-innerproduct}
\end{equation}
%

Note that because $u^{\rm TT}_{t\mu}\approx u^{\rm TT}_{r\mu}\approx 0$, 
only the angular component $h_{IJ}$ ($I,J=\theta,\phi$) appear 
in the right-hand side of Eq.~\eqref{Eq:integral-relation2}. 
From this, it follows that the right-hand side is invariant 
under the gauge transformation 
$h_{IJ}\rightarrow h_{IJ}+ \delta h_{IJ}$ with 
$\delta h_{IJ} = \nabla_{I} \xi_{J} + \nabla_{J}\xi_{I}$. 
This implies that $h_{\mu\nu}^{\rm}$  can be replaced 
by its counterpart in the TT gauge, $h^{\rm TT}_{\mu\nu}$. 
Hence, substituting the expansion formula~\eqref{expand-hmunu-scriplus} 
in the TT gauge into right-hand side of Eq.~\eqref{Eq:integral-relation2}, 
we obtain
%
\begin{equation}
C_{\tilde{j}}=8\pi i
\frac{
\left\langle {u}^{(\tilde{j})\mathrm{TT}},T
\right\rangle}
{\tilde{\omega} J_{\tilde{j}}}.
\label{relation-C-J}
\end{equation}
%
Here, we used the fact that $\hat{u}_{\mu\nu}^{(\tilde{j})\mathrm{TT}}$
behaves as $\sim e^{-i\tilde{\omega}(t-r)}/r$ for large $r$.

Furthermore, in spite of the fact that 
the relation~\eqref{Eq:integral-relation2}
was derived assuming the TT gauge conditions,
the inner product 
$\langle {u}^{(\tilde{j})\mathrm{TT}},{T}\rangle$ is invariant under the
gauge transformation $\hat{u}_{\mu\nu}^{\mathrm{TT}}\to 
\hat{u}_{\mu\nu}= \hat{u}_{\mu\nu}^{\mathrm{TT}}+\delta\hat{u}_{\mu\nu}$
with $\delta\hat{u}_{\mu\nu} = \nabla_{\mu}\xi_{\nu} + \nabla_{\nu}\xi_{\mu}$
because $\langle\delta u,T\rangle=0$ holds.
This can be shown by rewriting the integral of 
$(\nabla_{\mu}\xi_{\nu})T^{\mu\nu} = \nabla_{\mu}(T^{\mu\nu}\xi_{\nu})$
in the domain $D$ 
into the surface terms 
by the Gauss law and using the absence
of the energy flux at the boundaries $r=r_{\rm in}$ and $r_{\rm out}$. 
Therefore, the inner product 
$\langle \hat{u}^{(\tilde{j})},{T}\rangle$ can be calculated 
without taking a special care to the gauge choice, and
it is sufficient to find a solution 
$u^{(\tilde{j})}_{\mu\nu}$ satisfying the
TT gauge condition only at future null infinity.
In the Kerr background, we can explicitly construct such 
out-mode solutions as we will see later.

Substituting Eq.~\eqref{relation-C-J} in an arbitrary gauge
into Eq.~\eqref{PGW:general}, 
the energy emission rate is given by
%
\begin{equation}
\left\langle \frac{dE^{\rm (out)}_{\rm GW}}{dt} \right \rangle
  = \pi \sum_{\tilde{j}}
 \frac{\left|\langle u^{(\tilde{j})},T\rangle\right|^2}
{J_{\tilde{j}}}.
 \label{PGW:final}
\end{equation}
%
with $J_{\tilde{j}}$ defined in Eq.~\eqref{def:J_tildej}.

\subsection{Methods for calculating the energy emission rate}

In order to evaluate the radiation rate
using Eq.~\eqref{PGW:final},
we have to calculate the homogeneous solution $\hat{u}_{\mu\nu}$ 
for a gravitational perturbation and perform the integration 
of the inner product
$\langle u, T\rangle$. Also, the values of $J_{\tilde{j}}$
have to be calculated. 
We explain the methods for calculating these quantities.

\subsubsection{Homogeneous GWs}

We use the Teukolsky formalism \cite{Teukolsky:1973}
in order to calculate 
the homogeneous solution of a gravitational perturbation
of the Kerr spacetime. By setting $M=a=0$, this formalism
also can be applied to a perturbation for a flat spacetime
studied in the next section.
The Teukolsky formalism
realizes the separation of the variables $(t, r, \theta, \phi)$
of the perturbative equations using the Newman-Penrose
formalism \cite{Newman:1961}.
In the Teukolsky formalism, the Kinnersley null tetrads are adopted:
%
\begin{subequations}
\begin{eqnarray}
l^\mu &=& \frac{1}{\Delta}\left(r^2+a^2, \ 1, \ 0, \ a \right)
\\
n^\mu &=& \frac{1}{2\Sigma}\left(r^2+a^2, \ -\Delta, \ 0, \ a\right)
\\
m^\mu &=& \frac{1}{\sqrt{2}(r+ia\cos\theta)}
\left(ia\sin\theta, \ 0, \ 1, \ \frac{i}{\sin\theta}\right)
\end{eqnarray}
\end{subequations}
%
In terms of these tetrads, the metric~\eqref{Metric-Kerr-ST}
of the Kerr spacetime is 
%
\begin{equation}
g_{\mu\nu} = -l_\mu n_\nu - n_\mu l_\nu + m_\mu m^*_\nu + m_\nu m^*_\mu.
\end{equation}
%
The equations for the Newman-Penrose quantities $\psi_0$ and $\rho^{-4}\psi_4$
are given by
%
\begin{multline}
\left[\frac{(r^2+a^2)^2}{\Delta}-a^2\sin^2\theta\right]
\frac{\partial^2\psi}{\partial t^2}
+\frac{4Mar}{\Delta}\frac{\partial^2\psi}{\partial t\partial\phi}
+\left[\frac{a^2}{\Delta}-\frac{1}{\sin^2\theta}\right]
\frac{\partial^2\psi}{\partial\phi^2}
\\
-\Delta^{-s}\frac{\partial }{\partial r}
\left(\Delta^{s+1}\frac{d\psi}{dr}\right)
-\frac{1}{\sin\theta}\frac{\partial}{\partial\theta}
\left(\sin\theta\frac{\partial\psi}{\partial\theta}\right)
-2s\left[
\frac{a(r-M)}{\Delta}+\frac{i\cos\theta}{\sin^2\theta}
\right]\frac{\partial\psi}{\partial\phi}
\\
-2s\left[
\frac{M(r^2-a^2)}{\Delta}-r-ia\cos\theta
\right]
\frac{\partial\psi}{\partial t}
+(s^2\cot^2\theta-s)\psi = 0
\label{Teukolsky-master}
\end{multline}
%
with $s=\pm 2$, respectively. Here, the energy-momentum tensor
is set to be zero because we need to generate homogeneous solutions.
By writing $\psi = e^{-i\tilde{\omega}t}e^{i\tilde{m}\phi}
{}_sR_{\tilde{\ell}\tilde{m}}^{~\tilde{\omega}}(r)
{}_sS_{\tilde{\ell}\tilde{m}}^{~\tilde{\omega}}(\theta)$,
the Teukolsky equation~\eqref{Teukolsky-master} is separated as
%
\begin{equation}
\Delta^{-s}\frac{d}{dr}\left(\Delta^{s+1}
\frac{d{}_sR_{\tilde{\ell}\tilde{m}}^{~\tilde{\omega}}}{dr}\right)
+\left[\frac{\tilde{K}^2-2is(r-M)\tilde{K}}{\Delta}+4is\tilde{\omega} r-\lambda\right]
{}_sR_{\tilde{\ell}\tilde{m}}^{~\tilde{\omega}}=0,
\label{Eq:radial_Teukolsky_spin}
\end{equation}
%
%
\begin{multline}
\frac{1}{\sin\theta}\frac{d}{d\theta}
\left(\sin\theta\frac{d{}_sS_{\tilde{\ell}\tilde{m}}^{~\tilde{\omega}}}{d\theta}
\right)
\\
+\left(a^2\tilde{\omega}^2\cos^2\theta
-\frac{\tilde{m}^2}{\sin^2\theta}
-2a\tilde{\omega} s\cos\theta
-\frac{2\tilde{m}s\cos\theta}{\sin^2\theta}
-s^2\cot^2\theta
+s+{}_sA_{\tilde{\ell} \tilde{m}}
\right){}_sS_{\tilde{\ell}\tilde{m}}^{~\tilde{\omega}}=0,
\label{Eq:angular_Teukolsky_spin}
\end{multline}
%
where
%
\begin{equation}
\tilde{K}=(r^2+a^2)\tilde{\omega} - a\tilde{m}, \qquad
\lambda = {}_{s}A_{\tilde{\ell}\tilde{m}}+a^2\tilde{\omega}^2-2a\tilde{m}\tilde{\omega}.
\end{equation}
%

The function 
${}_sS_{\tilde{\ell}\tilde{m}}^{~\tilde{\omega}}(\theta)e^{i\tilde{m}\phi}$ 
is called a spin-weighted spheroidal harmonics,
and ${}_sA_{\tilde{\ell}\tilde{m}}$ is its eigenvalue.
For a spherically symmetric spacetime, the spin-weighted
spheroidal harmonics reduce to the spin-weighted spherical
harmonics ${}_sY_{\tilde{\ell}\tilde{m}}(\theta,\phi)$ with 
${}_sA_{\tilde{\ell}\tilde{m}}=\tilde{\ell}(\tilde{\ell}+1)-s(s+1)$. 
We adopt the standard normalization condition 
$\int \left|{}_sS_{\tilde{\ell}\tilde{m}}^{~\tilde{\omega}}(\theta)\right|^2
\sin\theta d\theta = 1/2\pi$.
For a Kerr spacetime, we have to generate 
${}_sS_{\tilde{\ell}\tilde{m}}^{~\tilde{\omega}}(\theta)$
and ${}_sA_{\tilde{\ell}\tilde{m}}$ by numerical calculation
or by approximate formulas. The method is explained in Sec.~\ref{Sec:IV2}.

As explained in Sec.~\ref{Sec:II2}, we calculate only 
the GW radiation rate toward infinity $dE_{\rm GW}^{\rm (out)}/dt$
that gives the upper bound on the total radiation rate.
In order to evaluate this quantity, we need to generate 
the out-mode solution $\hat{u}_{\mu\nu}$ for a gravitational perturbation of
a Kerr spacetime. The asymptotic behavior
of $R$ for the out-mode is
%
\begin{equation}
{}_sR_{\tilde{\ell}\tilde{m}}^{~\tilde{\omega}}\sim
\begin{cases}
\displaystyle
e^{i(\tilde{\omega}-\tilde{m}\Omega_H)r_*} & (r_*\to -\infty),\\
\displaystyle
Z_{\rm in}\frac{e^{-i\tilde{\omega}r_*}}{r}
+ Z_{\rm out}\frac{e^{i\tilde{\omega} r_*}}{r^{2s+1}} & (r_*\to +\infty),
\end{cases}
\label{Teukolsky-RadialFunction-Asymptotic}
\end{equation}
%
where $r_*$ is the tortoise coordinate
defined in Eq.~\eqref{Def:tortoise_coordinate}.
In this paper, we choose $s=+2$ for a technical reason
that is explained in Sec.~\ref{Sec:IV2}. 

Once the functions ${}_{+2}R(r)$ and ${}_{\pm 2}S(\theta)$
are obtained, the next step
is to reconstruct the metric perturbation $\hat{u}_{\mu\nu}$
from the Teukolsky functions.
The function $\psi_0$ or $\psi_4$ is known to completely
determine the metric perturbation \cite{Wald:1973}.
Intuitively, this is because
$\psi_0$ or $\psi_4$ is a gauge-invariant quantity,
and its real and imaginary parts correspond to the two degrees
of freedom of GWs. The explicit formula was derived
by Cohen and Kegeles~\cite{Cohen:1974,Cohen:1975}
and Chrzanowski~\cite{Chrzanowski:1975} under some assumptions, 
and its complete proof
was given by Wald~\cite{Wald:1978}. Here, we adopt 
the following metric formula given by Chrzanowski
in the outgoing radiation gauge 
$\hat{u}_{\mu\nu}n^\nu = \hat{u}_{\nu}^{~\nu} =0$:\footnote{
The authors of Ref.~\cite{Dias:2009} derived 
an apparently different formula 
and claimed that Chrzanowski's formula includes typos. However,
we have checked that both of the two formulas lead to the same result.}
\begin{multline}
\hat{u}_{\mu\nu}(\tilde{\ell}, \tilde{m}, {\tilde{\omega}}, P) 
=\left(
-n_\mu n_\nu\mathcal{A}
-m_\mu m_\nu\mathcal{B}
+n_{(\mu}m_{\nu)}\mathcal{C}\right)
{}_{+2}R_{\tilde{\ell}\tilde{m}}^{~\tilde{\omega}}(r)
{}_{-2}S_{\tilde{\ell}\tilde{m}}^{~\tilde{\omega}}(\theta)
e^{i\tilde{m}\phi-i\tilde{\omega}t}
\\
+P\left(
-n_\mu n_\nu\mathcal{A}^*
-m^*_\mu m^*_\nu\mathcal{B}^*
+n_{(\mu}m^*_{\nu)}\mathcal{C}^*
\right)
{}_{+2}R_{\tilde{\ell}\tilde{m}}^{~\tilde{\omega}}(r)
{}_{+2}S_{\tilde{\ell}\tilde{m}}^{~\tilde{\omega}}(\theta)
e^{i\tilde{m}\phi-i\tilde{\omega}t}
\label{Chrzanowski-formula}
\end{multline}
where $\mathcal{A}$, $\mathcal{B}$, and $\mathcal{C}$ are operators
\begin{subequations}
\begin{eqnarray}
\mathcal{A} &=&
\rho^{* -4}(\delta-3\alpha^*-\beta+5\pi^*)(\delta - 4\alpha^* + \pi^*),
\\
\mathcal{B} &=&
\rho^{* -4}(\Delta + 5\mu^* - 3\gamma^* + \gamma)(\Delta + \mu^*-4\gamma^*),
\\
\mathcal{C} &=&
\rho^{* -4}\left[
(\delta + 5\pi^*+\beta - 3\alpha^* + \tau)(\Delta + \mu^*-4\gamma^*)
\right.
\nonumber\\
&&
\left.
~~~
+(\Delta + 5\mu^*-\mu-3\gamma^*-\gamma)(\delta - 4\alpha^*+\pi^*)
\right].
\end{eqnarray}
\end{subequations}
Here, $\alpha$, $\beta$, $\gamma$, $\mu$, and $\pi$ are the Newman-Penrose
variables \cite{Newman:1961} (see \cite{Teukolsky:1973}
for expressions in the Kinnersley tetrad),
and $\Delta = n^\mu\nabla_\mu$, $\delta = m^\mu\nabla_\mu$.
The polarization-state parameter $P$ takes a value $+1$ or $-1$,
and the perturbations with $P=\pm 1$ are reduced 
to the even-type and odd-type perturbations of a Schwarzschild 
spacetime in the nonrotating case, $a_*=0$, 
respectively~\cite{Regge:1957,Zerilli:1970}
(or the scalar and vector modes in \cite{Kodama:2003}). 
For outgoing GWs, the outgoing radiation gauge
agrees with the TT gauge at the distant region $r\gg M$:
%
\begin{equation}
\hat{u}_{IJ} \approx \frac{Z_{\rm out}}{2}\tilde{\omega}^2
\frac{e^{-i\tilde{\omega} (t-r)}}{r}
\times
\begin{cases}
r^2\mathbb{S}_{IJ} & (P=+1),\\
r^2\mathbb{V}_{IJ} & (P=-1).
\end{cases}
\end{equation}
%
Here, we wrote only the $I,J=\theta, \phi$ components because the other
components are subdominant. 
The explicit formulas for $\mathbb{S}_{IJ}$ and $\mathbb{V}_{IJ}$
are
%
\begin{subequations}
\begin{eqnarray}
\mathbb{S}_{IJ} &=&
e^{i\tilde{m}\phi}\left(
\begin{array}{cc}
{}_{-2}S_{\tilde{\ell}\tilde{m}}^{~\tilde{\omega}}
+{}_{+2}S_{\tilde{\ell}\tilde{m}}^{~\tilde{\omega}} 
& 
i\sin\theta ({}_{-2}S_{\tilde{\ell}\tilde{m}}^{~\tilde{\omega}}
-{}_{+2}S_{\tilde{\ell}\tilde{m}}^{~\tilde{\omega}}) 
\\
* 
&
-\sin^2\theta ({}_{-2}S_{\tilde{\ell}\tilde{m}}^{~\tilde{\omega}}
+{}_{+2}S_{\tilde{\ell}\tilde{m}}^{~\tilde{\omega}})
\end{array}
\right),
\\
\mathbb{V}_{IJ} &=&
e^{i\tilde{m}\phi}\left(
\begin{array}{cccc}
 {}_{-2}S_{\tilde{\ell}\tilde{m}}^{~\tilde{\omega}}
-{}_{+2}S_{\tilde{\ell}\tilde{m}}^{~\tilde{\omega}} 
& 
i\sin\theta ({}_{-2}S_{\tilde{\ell}\tilde{m}}^{~\tilde{\omega}}
+{}_{+2}S_{\tilde{\ell}\tilde{m}}^{~\tilde{\omega}}) 
\\
 * 
& 
-\sin^2\theta ({}_{-2}S_{\tilde{\ell}\tilde{m}}^{~\tilde{\omega}}
-{}_{+2}S_{\tilde{\ell}\tilde{m}}^{~\tilde{\omega}})
\end{array}
\right).
\end{eqnarray}
\end{subequations}
%
Note that in the case of a spherically symmetric spacetime,
these tensors $\mathbb{S}_{IJ}$ and $\mathbb{V}_{IJ}$ are identical
to those given in Ref.~\cite{Kodama:2003} for the transverse-traceless
part of gravitational perturbation except for overall factors.
For this asymptotic form of $\hat{u}_{\mu\nu}$,
the value of $J$ defined in Eq.~\eqref{def:J_tildej} is calculated as
%
\begin{equation}
J = 2\left|Z_{\rm out}\right|^2\tilde{\omega}^4
\label{J-Chrzanowski}
\end{equation}
%
for both $P=\pm 1$.
Substituting Eq.~\eqref{J-Chrzanowski} into the
formula for the radiation rate~\eqref{PGW:final}, we have
%
\begin{equation}
\frac{dE_{\rm GW}}{dt} =
\frac{\pi}{2}
\left|\frac{\langle u, T\rangle}{Z_{\rm out}}\right|^2
\tilde{\omega}^{-4}.
\label{RadiationEfficiency_Zout}
\end{equation}
%

\subsubsection{Energy-momentum tensor}

Once the out-mode GW solution $u_{\mu\nu}$ and the value of $Z_{\rm out}$
are obtained, we can calculate the inner product using 
the energy-momentum tensor~\eqref{Energy-Momentum-Tensor-Original}.
Because the homogeneous solution $\hat{u}_{\mu \nu }$ of
Eq.~\eqref{Chrzanowski-formula} is spanned only by
$n_\mu n_\nu $, $n_{(\mu }m_{\nu )}$, $n_{(\mu }m^*_{\nu )}$,
$m_\mu m_\nu $, and $m^*_\mu m^*_\nu $,
the terms proportional to $g_{\mu\nu}$ do not contribute
to generation of GWs. For this reason, we only consider
$\hat{T}^{\prime}_{\mu\nu} = (1/2)\nabla_{\mu}\hat{\Phi}\nabla_{\nu}\hat{\Phi}$,
and the components
necessary for our calculation are
%
\begin{subequations}
\begin{eqnarray}
\hat{T}^{\prime\mu\nu}n_{\mu}n_{\nu} &=&
\frac{e^{-2i\omega t}e^{2im\phi}}{8\Sigma^2}
\left[iKR+\Delta R_{,r}\right]^2
S^2,
\label{EnergyMomentum-nn}
\\
\hat{T}^{\prime\mu\nu}m_{\mu}m_{\nu} &=&
\frac{e^{-2i\omega t}e^{2im\phi}}{4(r+ia\cos\theta)^2}
\left[S_{,\theta}+\left(a\omega\sin\theta-\frac{m}{\sin\theta}\right)S\right]^2
R^2,
\\
\hat{T}^{\prime\mu\nu}m^{*}_{\mu}m^{*}_{\nu} &=&
\frac{e^{-2i\omega t}e^{2im\phi}}{4(r-ia\cos\theta)^2}
\left[S_{,\theta}-\left(a\omega\sin\theta-\frac{m}{\sin\theta}\right)S\right]^2
R^2,
\\
\hat{T}^{\prime\mu\nu}n_{\mu}m_{\nu} &=&
\frac{-e^{-2i\omega t}e^{2im\phi}}{4\sqrt{2}\Sigma (r+ia\cos\theta)}
\left[iKR+\Delta R_{,r}\right]
\left[S_{,\theta}+\left(a\omega\sin\theta-\frac{m}{\sin\theta}\right)S\right]
SR,~~~~~
\\
\hat{T}^{\prime\mu\nu}n_{\mu}m^{*}_{\nu} &=&
\frac{-e^{-2i\omega t}e^{2im\phi}}{4\sqrt{2}\Sigma (r-ia\cos\theta)}
\left[iKR+\Delta R_{,r}\right]
\left[S_{,\theta}-\left(a\omega\sin\theta-\frac{m}{\sin\theta}\right)S\right]
SR,~~~~~
\label{EnergyMomentum-nms}
\end{eqnarray}
\end{subequations}
%
with $K=(r^2+a^2)\omega - am$. 

In the following two sections,
we apply this formalism to the analytic approximation in the
flat background spacetime and 
fully numerical calculations in the Kerr background spacetime,
respectively.

%
%
\section{Gravitational radiation in the flat approximation}

\label{Sec:III}

In this section, we derive approximate formulas for the GW radiation
rate in
the case $M\mu\ll 1$. In this case, most of the axion cloud distributes
in a region far from the BH. Detweiler \cite{Detweiler:1980}
constructed an approximate solution by the MAE method
for this case. As stated
earlier, the outer solution agrees with the wave function of a hydrogen
atom. The solution is
%
\begin{equation}
\hat{\Phi} =
-\frac{\sqrt{2E_a}}{\omega}(2k)^{3/2}
\sqrt{\frac{(n-\ell-1)!}{2n(n+\ell)!}}
e^{-i\omega t}
e^{-kr}(2kr)^\ell L_{n-\ell-1}^{2\ell+1}(2kr)Y_{\ell m}(\theta,\phi).
\label{wavefunction-atom}
\end{equation}
%
Here, $L_{n-\ell-1}^{2\ell+1}(x)$ represents the
Laguerre polynomial, and $k$ is defined by
%
\begin{equation}
k=\sqrt{\mu^2-\omega^2}.
\label{Def:k}
\end{equation}
%
The value of $k$ is discretized as
%
\begin{equation}
k = \frac{M\mu^2}{n},
\end{equation}
%
where $n=1,2,...$ is the principal quantum number defined
by
%
\begin{equation}
n:=\ell + 1 + n_r
\end{equation}
%
with the radial quantum number $n_r=0,1,2,...$.
Note that the gravitational analogue of the Bohr radius
is given by $a_0=1/M\mu^2 = n/k$.
In Eq.~\eqref{wavefunction-atom},
we added the extra factor $\sqrt{2E_a}/\omega$
compared to the standard wavefunction of a hydrogen atom 
in order to normalize the axion field so that
%
\begin{equation}
E_a = \int T_{tt}r^2drd\Omega
\end{equation}
%
is satisfied, where $d\Omega = \sin\theta d\theta d\phi$. 
We ignore the near-horizon solution
because it is small and gives a minor contribution to the
generation of GWs. In this approximation, the axion field
is bounded by the Newton potential
$-M\mu/r$, and the spacetime is approximately flat.
For this reason, we approximate the background metric $g_{\mu\nu}$ as the
flat spacetime metric $\eta_{\mu\nu}$ in calculating the
homogeneous GW solution and the inner product $\langle u,T\rangle$. 
We call this approximation {\it the flat approximation}.
In this flat approximation, 
we replace $g_{\mu\nu}$ of the energy-momentum 
tensor~\eqref{Energy-Momentum-Tensor-Original} by $\eta_{\mu\nu}$, 
and raise and lower the tensor indices $\mu, \nu$
by $\eta_{\mu\nu}$. 

Here, we point out a potential problem in this approximate method.
The approximate solution~\eqref{wavefunction-atom} 
for the axion field satisfies the equation
\begin{equation}
\eta^{\mu\nu}\partial_\mu\partial_\nu\Phi 
- \mu^2\Phi = -\frac{2kn}{r}\Phi.
\end{equation}
The right-hand side comes from the Newton potential, or in
other words, the contribution from the static perturbation
generated by the BH mass $M$.
On the other hand, in calculating the homogeneous GWs and the
inner product, 
we completely ignore the background curvature and
use the flat metric $\eta_{\mu\nu}$. 
As a result, we have
\begin{equation}
\partial_\mu T^{\mu\nu} = -\frac{kn}{r}\partial^\nu\Phi^2,
\end{equation}
and the conservation of the energy and momentum
is weakly violated. Because of this, 
if we adopt the gauge transformation 
$\delta u_{\mu\nu}=\partial_\mu\xi_\nu+\partial_\nu\xi_\mu$,
we have
\begin{equation}
\langle \delta u, T\rangle
=
\frac{1}{\Delta t}\int\frac{kn}{r}\xi^\mu\partial_\mu\Phi^2\sqrt{-g}d^4x,
\label{InnerProduct-GaugeTransformation}
\end{equation}
and therefore, the gauge invariance
of the inner product is not guaranteed.
For this reason, we have to check carefully to what extent
the ``error'' in Eq.~\eqref{InnerProduct-GaugeTransformation}
can be large. We will come back to this point at the last
part of this section. 

\subsection{Solution to Teukolsky equation}

\label{Sec:III1}

In the flat spacetime, the radial Teukolsky equation becomes
\begin{equation}
\frac{d^2{}_{s}R_{\tilde{\ell}\tilde{m}}}{dx^2} 
+ \frac{2s+2}{x}\frac{d{}_{s}R_{\tilde{\ell}\tilde{m}}}{dx}
+\left[1+\frac{2is}{x}-\frac{(\tilde{\ell}-s)(\tilde{\ell}+s+1)}{x^2}\right]
{}_{s}R_{\tilde{\ell}\tilde{m}}
=0.
\end{equation}
Here, the rescaled radial coordinate $x=\tilde{\omega}r$ is
introduced. This equation can be solved analytically.
Choosing $s=+2$, the general solution is
given as 
\begin{equation}
{}_{+2}R_{\tilde{\ell}\tilde{m}} = 
A e^{-ix}x^{\tilde{\ell}-2}U(\tilde{\ell}-1,2\tilde{\ell}+2,2ix)
+B e^{-ix}x^{\tilde{\ell}-2}F(\tilde{\ell}-1,2\tilde{\ell}+2,2ix),
\end{equation}
where $F$ and $U$ denote the confluent hypergeometric functions
of the first and second kinds, respectively. 
We choose the integral constants to be $A=0$, $B=1$;
\begin{equation}
{}_{+2}R_{\tilde{\ell}\tilde{m}} 
= e^{-ix}x^{\tilde{\ell}-2}F(\tilde{\ell}-1,2\tilde{\ell}+2,2ix),
\label{Teukol-Solution}
\end{equation}
so that the radial function becomes regular at the origin, $x=0$.
The asymptotic behavior at large $x$ is
\begin{equation}
{}_{+2}R_{\tilde{\ell}\tilde{m}}\approx
\frac{\Gamma(2\tilde{\ell}+2)}{(2i)^{\tilde{\ell}+3}\Gamma(\tilde{\ell}-1)}
\frac{e^{ix}}{x^5}
+
\frac{\Gamma(2\tilde{\ell}+2)}{(-2i)^{\tilde{\ell}-1}\Gamma(\tilde{\ell}+3)}
\frac{e^{-ix}}{x}.
\end{equation}
Here, the first term represents outgoing waves.
From this asymptotic behavior, 
the coefficient $Z_{\rm out}$ 
for the asymptotic outgoing waves 
in Eq.~\eqref{Teukolsky-RadialFunction-Asymptotic}
is read to be
\begin{equation}
Z_{\rm out}
=
\frac{\Gamma(2\tilde{\ell}+2)}{(2i)^{\tilde{\ell}+3}\Gamma(\tilde{\ell}-1)}
\tilde{\omega}^{-5}.
\label{Zout}
\end{equation}


\subsection{Energy emission rate}
\label{Sec:III2}

The remaining task is to obtain the homogeneous solution
$\hat{u}^{(\tilde{j})}_{\mu\nu}$ and calculate the integral
$\langle u^{(\tilde{j})}, T\rangle$, where the index $(\tilde{j})$ 
is a shorthand for $(\tilde{\ell},\tilde{m}, P)$
to label each GW mode
as noted in Sec.~\ref{Sec:II2}.
Since the calculation of the inner product
is rather tedious, we sketch the 
calculations in Appendix~B, and
just present the results here.
It is convenient to 
discuss the results for the odd-type GWs ($P=-1$) and 
for the even-type GWs ($P=+1$) separately.


\subsubsection{Odd-type perturbation}

In the case of odd-type perturbation, $P=-1$,  
the inner product $\langle u,T\rangle$
vanishes after integration with respect to the 
angular coordinates $(\theta, \phi)$.
Therefore, we find that {\it GWs of the odd-type modes are not radiated}
in our BH-axion system in the flat approximation. This is a natural
result because the oscillating part of the energy-momentum tensor
of the axion cloud has just the even-type mode. The stationary part of the
energy-momentum tensor generates the odd-type perturbation
that corresponds to the gravitational angular momentum,
and this part has been ignored in our analysis.

\subsubsection{Even-type perturbation}

We discuss the case of even-type perturbation, $P=+1$. 
Here, we assume the axion cloud to be in the mode $\ell = m$
for simplicity. Then, 
it turns out that only GWs of the mode 
$\tilde{\ell}=\tilde{m}=2\ell$ are radiated,
and the energy radiation rate is given by
%
\begin{subequations}
\begin{equation}
\frac{dE_{\rm GW}}{dt} =
C_{n\ell} \left(\frac{E_a}{M}\right)^2(\mu M)^{Q_{\ell}},
\label{Flat_GW_efficiency_result}
\end{equation}
where
\begin{equation}
Q_{\ell} = 4\ell+10,
\label{QL}
\end{equation}
and
\begin{equation}
C_{n\ell} 
= 
\frac{16^{\ell+1}\ell(2\ell-1)\Gamma(2\ell-1)^2\Gamma(\ell+n+1)^2}
{n^{4\ell+8}(\ell+1)\Gamma(\ell+1)^4\Gamma(4\ell+3)\Gamma(n-\ell)^2}
\label{CNL}
\end{equation}
\end{subequations}
%
Table~\ref{Table:flat-approximation} summarizes the values
of $C_{n\ell}$ and $Q_{\ell}$ and the radiated GW mode
$(\tilde{\ell}, \tilde{m})$ for several axion modes.

%
\begin{table}[tb]
\caption{Approximate values of $C_{n\ell }$, values of
$Q_{\ell}$, and the emitted GW mode $(\tilde{\ell}, \tilde{m})$ 
for several axion cloud modes $(n, \ell, m)$. }
\centering
\begin{tabular}{ccccc}
\hline\hline
Axion mode $(n,\ell,m)$ & Atomic orbital & $C_{n\ell }$ & $Q_{\ell}$ & GW mode $(\tilde{\ell},\tilde{m})$\\
  \hline
$(2,1,1)$ & 2p & $1.56\times 10^{-3~}$ & $14$ & $(2,2)$\\
$(3,1,1)$ & 3p & $1.93\times 10^{-4~}$ & $14$ & $(2,2)$\\
$(4,1,1)$ & 4p & $3.81\times 10^{-5~}$ & $14$ & $(2,2)$\\
$(3,2,2)$ & 3d & $1.89\times 10^{-7~}$ & $18$ & $(4,4)$\\
$(4,2,2)$ & 4d & $6.81\times 10^{-8~}$ & $18$ & $(4,4)$\\
$(5,2,2)$ & 5d & $2.35\times 10^{-9~}$ & $18$ & $(4,4)$\\
$(4,3,3)$ & 4f & $2.89\times 10^{-11}$ & $22$ & $(6,6)$\\
$(5,3,3)$ & 5f & $2.14\times 10^{-11}$ & $22$ & $(6,6)$\\
$(6,3,3)$ & 6f & $1.13\times 10^{-11}$ & $22$ & $(6,6)$\\
$(5,4,4)$ & 5g & $3.17\times 10^{-15}$ & $26$ & $(8,8)$\\
$(6,4,4)$ & 6g & $3.98\times 10^{-15}$ & $26$ & $(8,8)$\\
$(7,4,4)$ & 7g & $2.97\times 10^{-15}$ & $26$ & $(8,8)$\\
\hline\hline
\end{tabular}
\label{Table:flat-approximation}
\end{table}
%

\subsection{Comparison with the superradiant growth rate}

Here, we compare the GW radiation rate with the
energy extraction rate of the axion cloud by the superradiant instability.
For the situation $M\mu\ll 1$, an approximate formula for 
the growth rate by the instability has been 
derived by Detweiler~\cite{Detweiler:1980}
with the MAE method. The approximate growth rate is given by
%
\begin{multline}
M\omega_I = (M\mu)^{4\ell+5}(ma_*-2\mu r_+)
\frac{2^{4\ell+2}(n+\ell)!}{n^{2\ell+4}(n-\ell-1)!}
\left[\frac{\ell!}{(2\ell+1)!(2\ell)!}\right]^2
\\
\times
\prod_{j=1}^{\ell}\left[(1-a_*^2)j^2+(ma_*-2\mu r_+)^2\right].
\end{multline}
%
Note that the definition of $n$ in Ref.~\cite{Detweiler:1980} is different
from ours: it corresponds to our radial quantum
number $n_r$. With this $\omega_I$, the energy extraction rate
by the superradiant instability is given by
Eq.~\eqref{superradiant-energy-extraction-rate}.
Because the energy extraction rate
and the GW radiation rate are proportional to
$(M\mu)^{4\ell + 5}$ and $(M\mu)^{Q_\ell}$ with $Q_\ell=4\ell+10$, 
respectively, the GW radiation rate
has a higher power of $(M\mu)$.
Since our approximation holds for $M\mu\ll 1$,
the GW radiation rate is expected
to be much smaller than the energy extraction.

%
\begin{figure}[tb]
\centering
\includegraphics[width=2.5in]{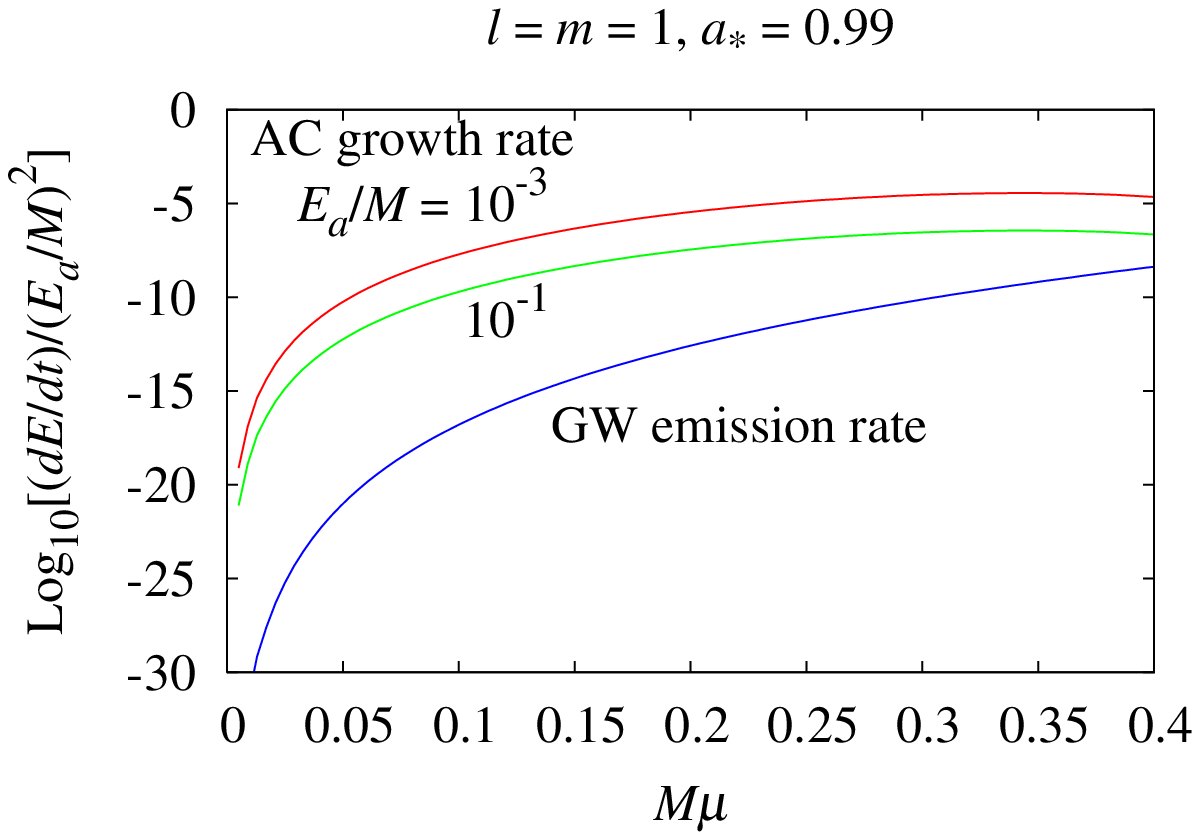}
\includegraphics[width=2.5in]{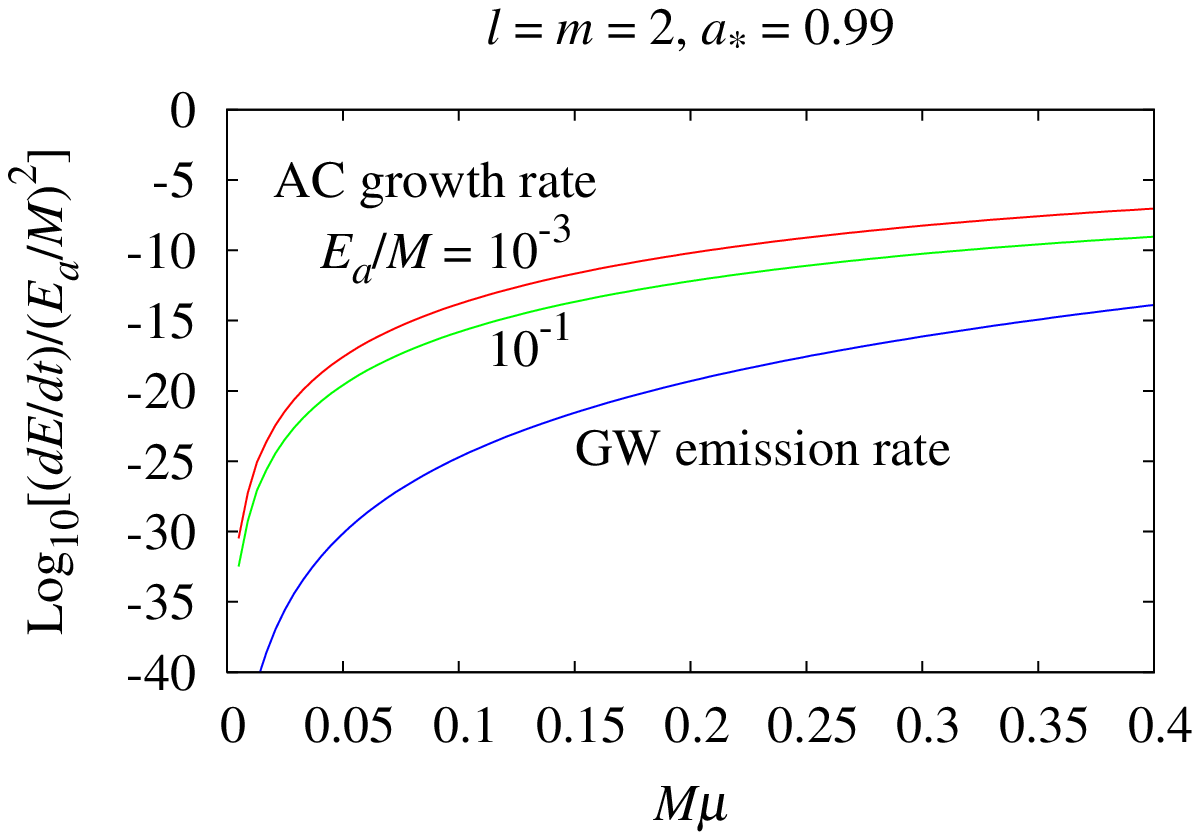}
\includegraphics[width=2.5in]{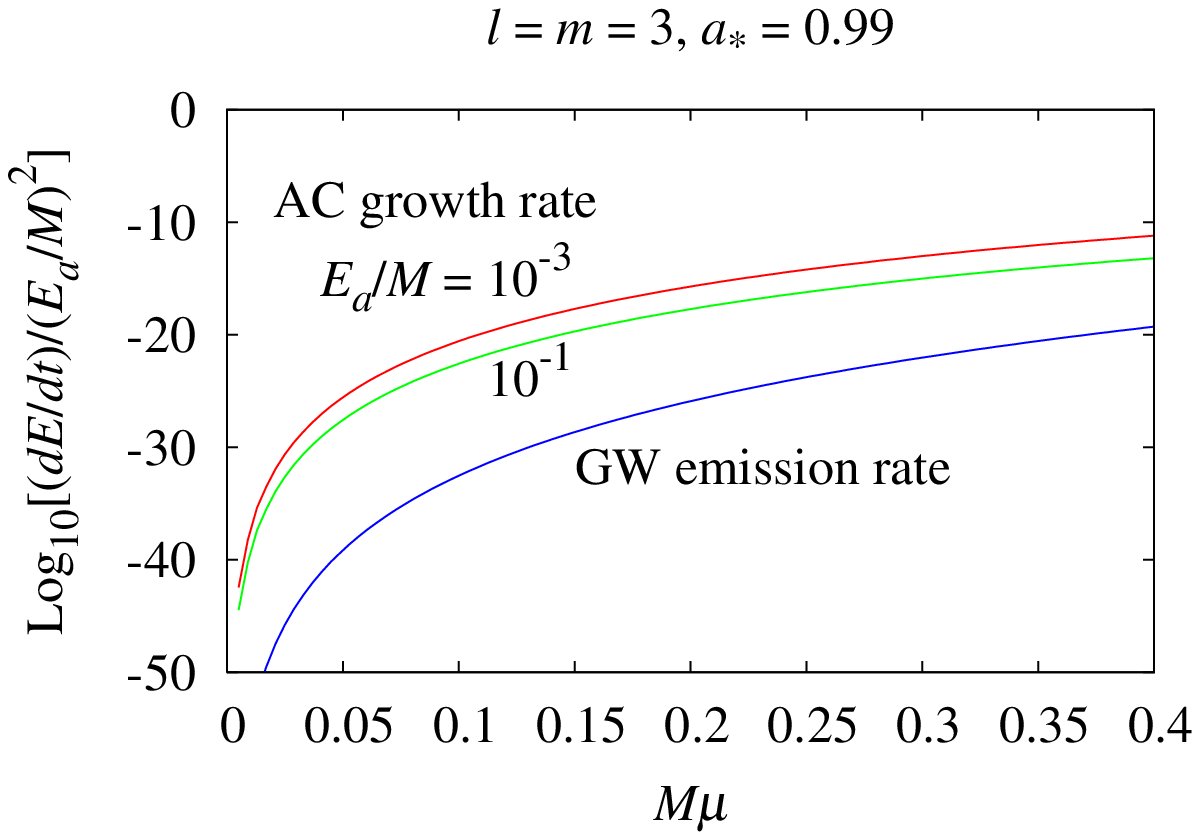}
\caption{The GW radiation rate 
normalized by $(E_a/M)^2$ as functions of $M\mu$ in the flat approximation.
The cases of the axion cloud in the mode
$\ell = m = 1$, $2$, and $3$, and $n_r=0$ are shown 
for the BH rotation parameter $a_*=0.99$.
The energy extraction rates of the axion cloud are also plotted
for the two cases $E_a/M=10^{-3}$ and
$10^{-1}$ for comparison. In all cases, the energy extraction rate is larger
than the GW radiation rate.}
\label{Flat_GW_AC}
\end{figure}
%

Figure~\ref{Flat_GW_AC} shows the GW radiation rate $dE_{\rm GW}/dt$
and the energy extraction rate $dE_a/dt$ by the superradiant instability
normalized by $(E_a/M)^2$ 
as functions of $M\mu$ for $\ell=m=1$, $2$, and $3$.
The BH rotation parameter is fixed to be $a_*=0.99$,
and we show the two cases where the energy of the axion cloud
is  $E_a/M\sim 10^{-3}$ and $10^{-1}$. These two values correspond
to the energies when the bosenova happens for the choice of the decay constant
$f_a=10^{16}$ and $10^{17}$ GeV, respectively \cite{Yoshino:2012}.
The figure shows that the energy extraction rate
is much larger. For other values of the BH
rotation parameter $a_*$, we have found that
the result does not change in the region
where the superradiant instability works effectively. Therefore,
in the region $M\mu\ll 1$, 
the GW radiation rate is smaller than the
energy extraction rate, and the GW emission 
does not hinder the growth of the axion cloud by 
the superradiant instability
for a wide range of system parameters.

\subsection{Reliability of the flat approximation}

\label{Sec:III4}

Now, we discuss the problem of the gauge dependence
of the inner product in the flat approximation, 
Eq.~\eqref{InnerProduct-GaugeTransformation}.
Here, we consider the gauge transformation $\delta u$ of $u$
generated by a vector field
$\xi^\mu \propto e^{-i\tilde{\omega}t}$. 
Since $\delta u\sim \omega\xi$ and $\partial_t\Phi^2\sim \omega \Phi^2$,
we have
\begin{equation}
\langle \delta u,T\rangle \sim \frac{\beta\omega^2}{\Delta t}
\int \delta u\frac{\Phi^2}{x}\sqrt{-g}d^4x.
\end{equation}
Here, we have introduced the small parameter $\beta:=k/\omega\approx \mu M/n$.
On the other hand, since the order of the $tt$ component
of the energy-momentum tensor is $T_{tt}\sim \omega^2 \Phi^2$, 
readers may expect that 
$\langle u,T\rangle \sim ({\omega^2}/{\Delta t})
\int u{\Phi^2}\sqrt{-g}d^4x$, and thus, 
$\langle \delta u,T\rangle \sim \beta \langle u,T\rangle
\ll \langle u, T\rangle$. However, this is incorrect
because the leading order terms with respect to $\beta$
cancel out in the calculation of $\langle u,T\rangle$
as explained just before Eq.~\eqref{InnerProduct2} in Appendix~B.
The fact is that because of this cancellation,
\begin{equation}
\langle u,T\rangle \sim \frac{\beta\omega^2}{\Delta t}
\int u\Phi^2\sqrt{-g}d^4x,
\end{equation} 
and therefore,  $\langle u, T\rangle\sim \langle\delta u, T\rangle$.
The change in the inner product caused by the gauge transformation
contribute to the leading order of $\langle u, T\rangle$.
Although this problem of the gauge dependence would be 
solved by introducing the
static perturbation from a flat spacetime that corresponds
to the Newton potential, the 
analytic calculation will become much more difficult
in such an analysis.

From the observations above, the value of the coefficients
$C_{n\ell}$ of Eq.~\eqref{CNL} is not reliable and may be changed
by a factor. In fact, we also calculated the inner product
$\langle u, T\rangle$ using the
gauge-invariant formalism for perturbations in spherically symmetric spacetimes
in general dimensions~\cite{Kodama:2003} in the
radiation gauge $u_{0\mu}=u^{\mu}_{~\mu}=0$ and found that 
the prefactor $C_{n\ell}$ for the radiation 
formula~\eqref{Flat_GW_efficiency_result} for the even-type mode 
differs from Eq.~\eqref{CNL} by a factor of $(\ell-1)^2/4$ 
(although the result for the odd-type mode is unchanged).
However, we would like to stress that the power 
dependence on $(\mu M)$ must not be changed by the gauge choice:
{\it We can trust the power $Q_{\ell}$ of $(\mu M)$ given by Eq.~\eqref{QL}.}
We will confirm this statement in the fully numerical calculation
in the Kerr background spacetime in the next section.

Note that this gauge dependence 
just appears in the case of the flat approximation. In the full numerical
calculations in the Kerr background below, the gauge invariance
of the inner product is guaranteed because the energy conservation
$\nabla_\mu T^{\mu\nu}=0$ is fully satisfied.

%
%
\section{Gravitational radiation in Kerr background}

\label{Sec:IV}

In this section, we explain
the numerical method of computing the GW radiation rate from an axion cloud
in the Kerr background spacetime and present the numerical results.
This analysis can be applied for the range of the system parameter $M\mu\sim 1$.

\subsection{Numerical method}
\label{Sec:IV2}

The numerical method of calculating the axion bound state
$\hat{\Phi}=e^{-i\omega t}e^{im\phi}R(r)S(\theta)$
was already explained in Sec.~2.1. The bound-state frequency
and the radial function $R(r)$
are obtained by the continued fraction method.
The angular function $S(\theta)$ is approximately calculated
with an expansion formula $S(\theta) = \sum_i S^{(i)}c^{i}$ with
$c^2 = -a^2k^2$ up to the sixth order
for each $(\ell, m)$. See Appendix~\ref{Appendix-A} for details.
Once these functions are obtained, the necessary components
of the energy-momentum tensor are calculated with 
Eqs.~\eqref{EnergyMomentum-nn}--\eqref{EnergyMomentum-nms}.

In order to generate the homogeneous solutions $u^{(\tilde{j})}_{\mu\nu}$, 
we first calculate the functions and quantities
that appear in the Teukolsky function: 
The spin-weighted spheroidal harmonics 
${}_sS_{\tilde{\ell}\tilde{m}}^{~\tilde{\omega}}(\theta)e^{i\tilde{m}\phi}$ 
and its eigenvalue ${}_sA_{\tilde{\ell}\tilde{m}}^{~\tilde{\omega}}$, and 
the radial function ${}_sR_{\tilde{\ell}\tilde{m}}^{~\tilde{\omega}}(r)$.

For the eigenvalue ${}_sA_{\tilde{\ell}\tilde{m}}^{~\tilde{\omega}}$, 
we adopt
the approximate formula 
for small $c:=a\tilde{\omega}$ given
in Refs.~\cite{Seidel:1988,Berti:2005} that is applicable for arbitrary 
values of $(s, \tilde{\ell}, \tilde{m})$. 
This approximate formula is given in the form of the
series expansion with respect to $c=a\tilde{\omega}$ up to the sixth order.
Since approximate formulas for the functions 
${}_sS_{\tilde{\ell}\tilde{m}}^{~\tilde{\omega}}$ have not been given in
an existing literature, we have derived the series expansion
${}_sS(\theta) = \sum_{i}{}_sS^{(i)}c^i$ up to the sixth order
for each value of $(s, \tilde{\ell}, \tilde{m})$. 
The reason why we use the approximate formula is that regularization
at poles is necessary when we calculate the metric from the Teukolsky
functions. This regularization procedure is much more
difficult for numerically generated data.

The radial function ${}_{+2}R_{\tilde{\ell}\tilde{m}}^{~\tilde{\omega}}(r_*)$ 
was numerically generated using the fourth-order
Runge-Kutta method starting from $r_*=-200$ to outward
up to $r_*/M = 1000$ or $4000$ depending on the situation.
Here, we choose $s=+2$
because the ingoing mode behaves as
${}_{s}R_{\tilde{\ell}\tilde{m}}^{~\tilde{\omega}}
\sim \Delta^{-s}e^{-i(\tilde{\omega}-\tilde{m}\Omega_H)r}$ 
near the horizon and
it is easy to kill this mode for a positive $s$. 
In order to calculate the radiation rate, we have to
determine the value of $Z_{\rm out}$ from the numerical data.
This is done by fitting
the numerical data with the asymptotic expansion formula
of the form
%
\begin{equation}
{}_{+2}R_{\tilde{\ell}\tilde{m}}^{~\tilde{\omega}}
\approx
Z_{\rm in}\frac{e^{-i\tilde{\omega} r}}{r^{1+2M\tilde{\omega} i}}
\left[1 + \frac{A_1}{r} + \cdots + \frac{A_5}{r^5}  \right]
\\
+Z_{\rm out}\frac{e^{i\tilde{\omega} r}}{r^{5-2M\tilde{\omega} i}}
\left[1 + \frac{B_1}{r} + \frac{B_2}{r^2} \right].
\end{equation}
%
Here, the coefficients $A_1,...,A_5$ and $B_1, B_2$ are
determined from the asymptotic expansion of
the radial Teukolsky equation~\eqref{Eq:radial_Teukolsky_spin},
and the fitting parameters are $Z_{\rm in}$ and $Z_{\rm out}$.
Although the outgoing mode decays more rapidly compared to the ingoing mode
at large $r$, 
the value of $Z_{\rm out}$ can be evaluated fairly accurately
by increasing the numerical accuracy of 
${}_{+2}R_{\tilde{\ell}\tilde{m}}^{~\tilde{\omega}}(r)$.
The fit was done in the range $100\le r_*/M\le 300$, because
the numerical noise increases as the value of $r_*/M$ is further increased.

\subsection{General properties}

The modes of radiated GWs can be restricted
by the symmetry properties of the (spin-weighted) spheroidal harmonics.
Assuming that the axion cloud is in the mode
$\ell = m$, we have
%
\begin{equation}
S_{\ell m}(\theta) = S_{\ell m}(\pi-\theta).
\end{equation}
%
On the other hand, the spin-weighted spheroidal harmonics
for the even azimuthal quantum number $\tilde{m}=2m$ satisfies
%
\begin{equation}
{}_{+2}S_{\tilde{\ell} \tilde{m}}^{~\tilde{\omega}}(\theta) =
(-1)^{\tilde{\ell}} {}_{-2}S_{\tilde{\ell} \tilde{m}}^{~\tilde{\omega}}(\pi-\theta).
\end{equation}
%
From these properties, the angular integration
of $\langle u,T\rangle$ becomes zero 
when 
$P=(-1)^{\tilde{\ell}+1}$. Therefore,
the radiated GW modes are limited to the ones satisfying
$P=(-1)^{\tilde{\ell}}$. For the ``even-type'' perturbations
$(P=+1)$, the GW modes with the quantum numbers
$\tilde{\ell}-\tilde{m}=0, 2, 4,...$ 
are radiated, and for the ``odd-type'' perturbations
$(P=-1)$, those with the quantum numbers 
$\tilde{\ell}-\tilde{m} =1, 3, 5,...$ are radiated.

In the flat approximation in the previous section,
all vector modes ($P=-1$) vanished. 
But in the case of the rotating Kerr background,
our numerical data show that the $P=-1$ modes are nonzero
as we see below. The reason can be understood by considering
the slow rotation case. The effect of the BH rotation  
is given by the odd-type perturbation, and the coupling
between the even-type axion distribution 
and the odd-type rotation becomes the source for the  
odd-type GWs at the second order.

\subsection{Numerical results}

Now we show the numerical results.

\subsubsection{$\ell=m=1$}

%
\begin{figure}[tb]
\centering
\includegraphics[width=2.5in]{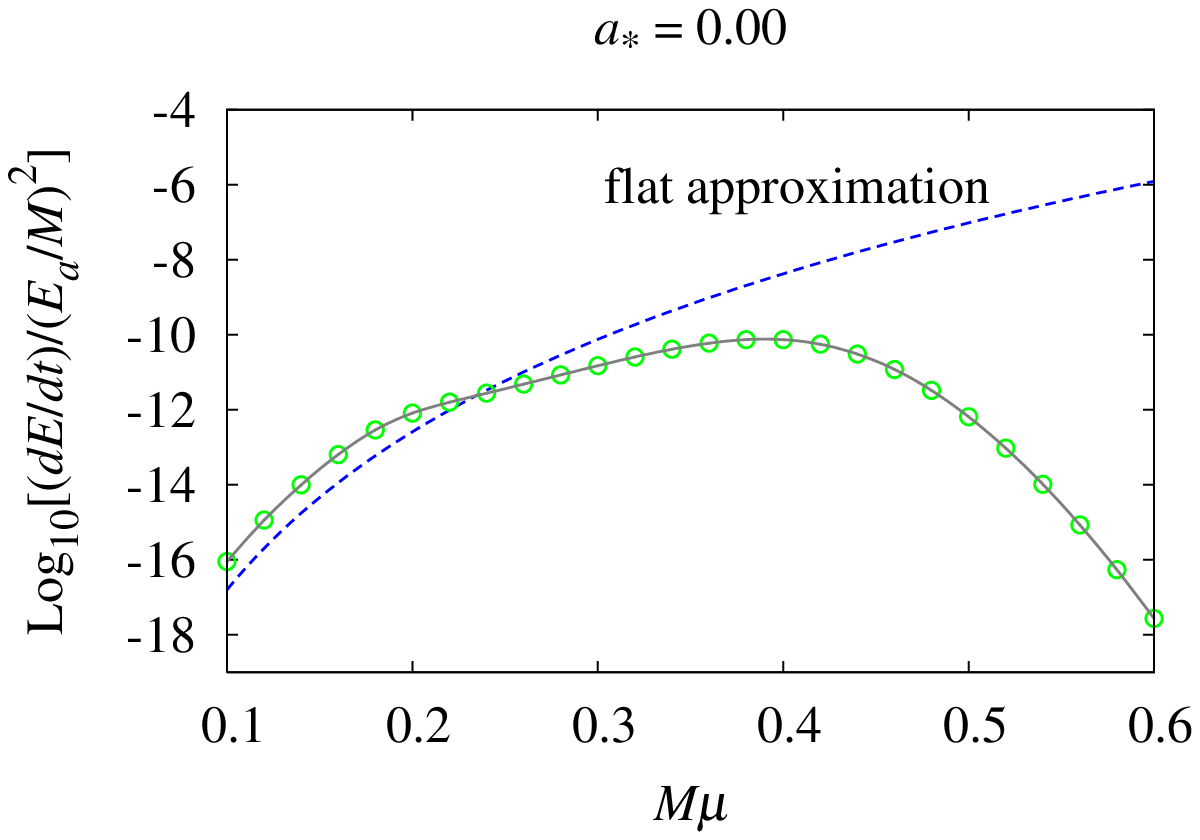}
\includegraphics[width=2.5in]{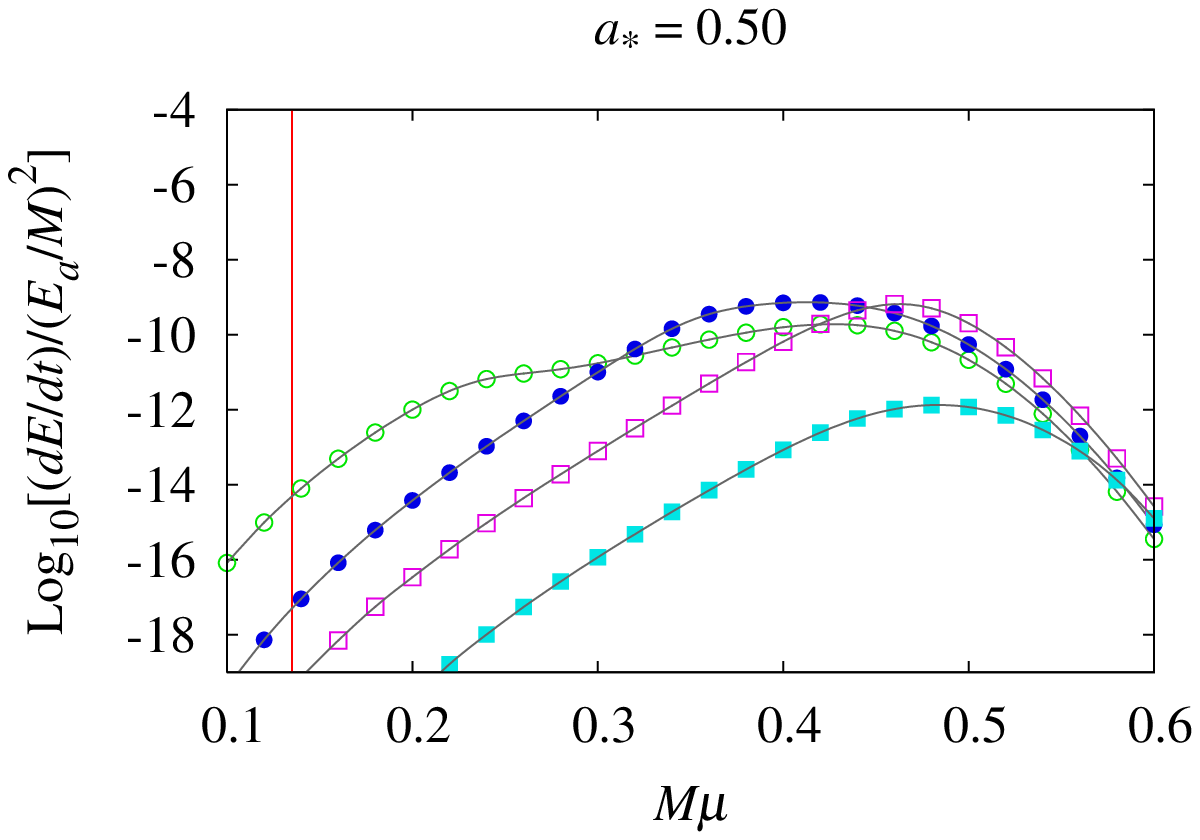}
\includegraphics[width=2.5in]{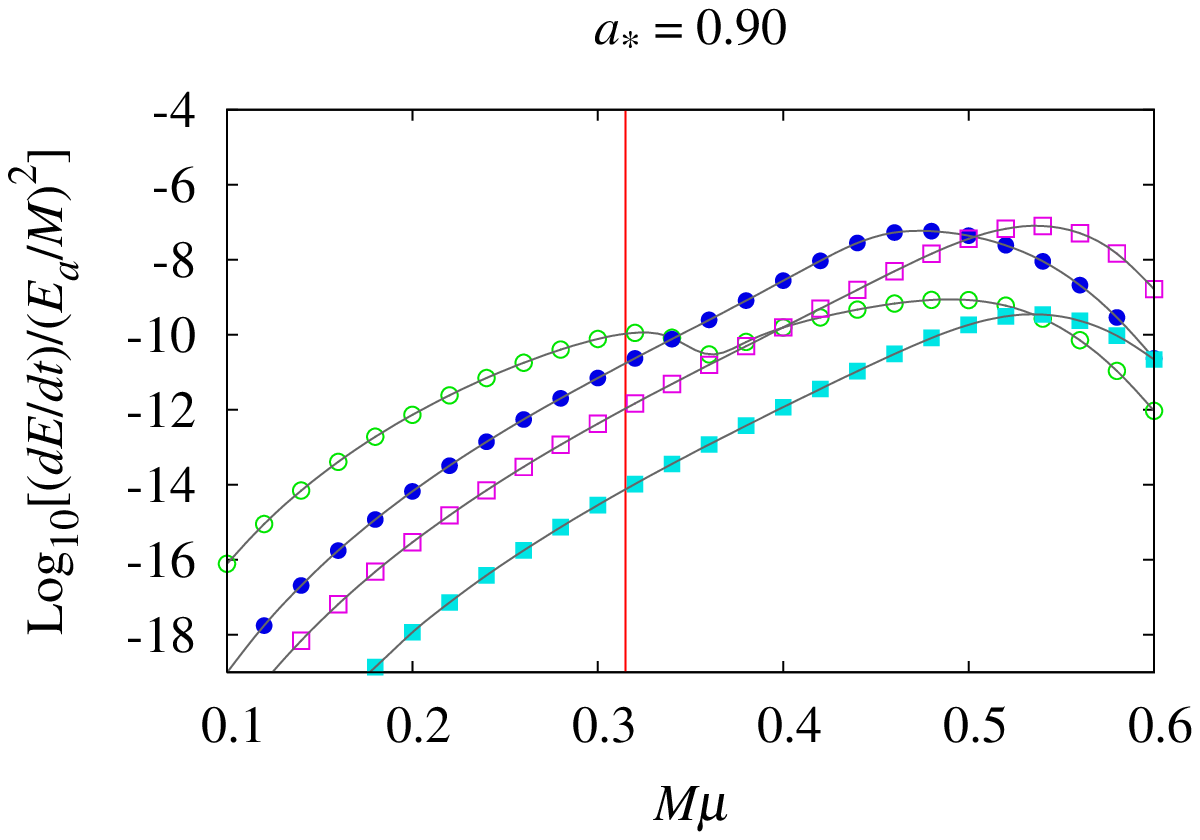}
\includegraphics[width=2.5in]{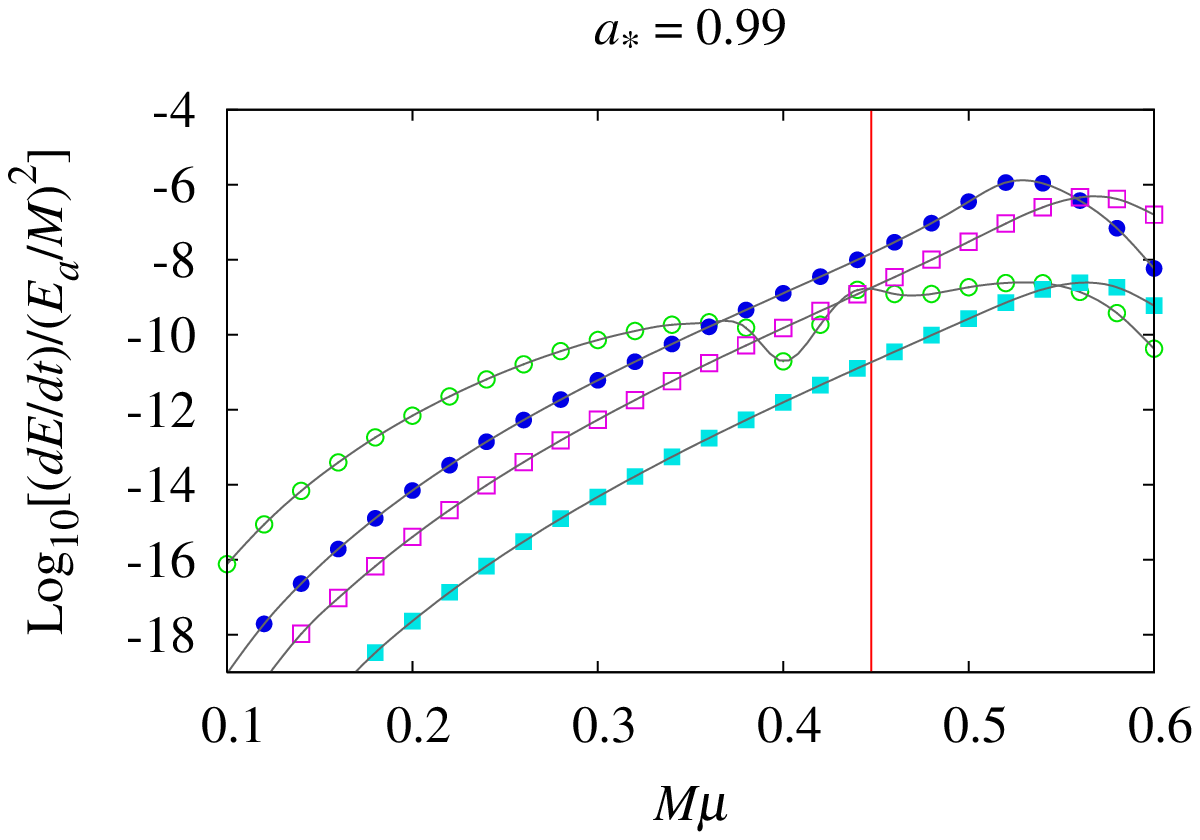}
\caption{GW energy radiation rates 
for the modes $(\tilde{\ell}, \tilde{m}) = (2,2)$ ($\odot$),
$(3,2)$ ($\bullet$),
$(4,2)$ ($\square$),
and $(5,2)$ ($\blacksquare$) from the axion cloud 
in the $(\ell, m) = (1,1)$ mode, as functions of $M\mu$
in the cases of $a_*=0.00$,
$0.50$, $0.90$, and $0.99$. The result for the flat approximation
is shown by a dashed curve in the figure of $a_*=0.00$ and
the vertical lines indicate the border of the superradiant instability
for each $a_*$.}
\label{GW_rate_mode_L1M1}
\end{figure}
%

We begin with the results for the axion cloud in the $\ell = m = 1$
mode that is reduced to the 
gravitational atom with 
a vanishing radial quantum number $n_r=0$ for $M\mu\ll 1$.
Figure~\ref{GW_rate_mode_L1M1} shows the GW radiation rates
$dE_{\rm GW}^{(\tilde{\ell}\tilde{m})}/dt$
normalized by $(E_a/M)^2$ for the modes
$(\tilde{\ell}, \tilde{m})=(2,2)$ (circles, $\odot$),
$(3,2)$ (black circles, $\bullet$),
$(4,2)$ (squares, $\square$), and
$(5,2)$ (black squares, $\blacksquare$)
as functions of $M\mu$
for $a_*=0.00$, $0.50$, $0.90$, and $0.99$. The value of
$P$ for each GW mode is $P=(-1)^{\tilde{\ell}}$.
In the panel of $a_*=0.00$, only the mode $(\tilde{\ell}, \tilde{m})=(2,2)$
is shown because other modes become zero with the same reason
as the flat approximation. However,
as the value of $a_*$ is increased, the contribution
of other modes becomes important. In fact, there are regions where
the modes $(\tilde{\ell}, \tilde{m})=(3,2)$ or $(4,2)$
become dominant.

In each panel of $a_*=0.50$, $0.90$, and $0.99$, the vertical line
indicates the threshold of the superradiant instability: In the left-hand
side of the line, the axion cloud grows by superradiant instability at least if the GW emission is neglected. 
In the right-hand side, no superradiant instability occurs, 
and the axion cloud, even if it were produced by some mechanism, 
would simply shrink gradually due to infall into the BH.
We find the general tendency such that the radiation rate
decreases for very large $M\mu$. The reason is as follows.
In the superradiant regime, there is always a potential minimum
of the axion field
and the quasibound state is formed. Although this potential minimum
also exists for $M\mu$ that is not much larger than
the threshold, it disappears at some point as $M\mu$ is further increased.
In such a situation, the field cannot form a bound state
and it just falls into the horizon. 
Then, the field is concentrated near the horizon 
and the GW emission becomes inefficient, 
because the redshift effect becomes more and more significant as 
$M\mu$ increases further. 
The GW radiation rate decreases rapidly with $M\mu$.

In the panel of $a_*=0.00$, the result of the flat approximation
is also shown. For a small value of $M\mu$, 
our numerical data obey the same 
power dependence on $M\mu$ as the approximate formula,
$dE_{GW}/dt\propto (M\mu)^{14}$. 
There is a shift by a constant factor from the curve
of the approximate formula, because the value of 
$C_{n\ell}$ includes the error by a factor as discussed 
in Sec.~\ref{Sec:III4}.

%
\begin{figure}[tb]
\centering
\includegraphics[width=2.5in]{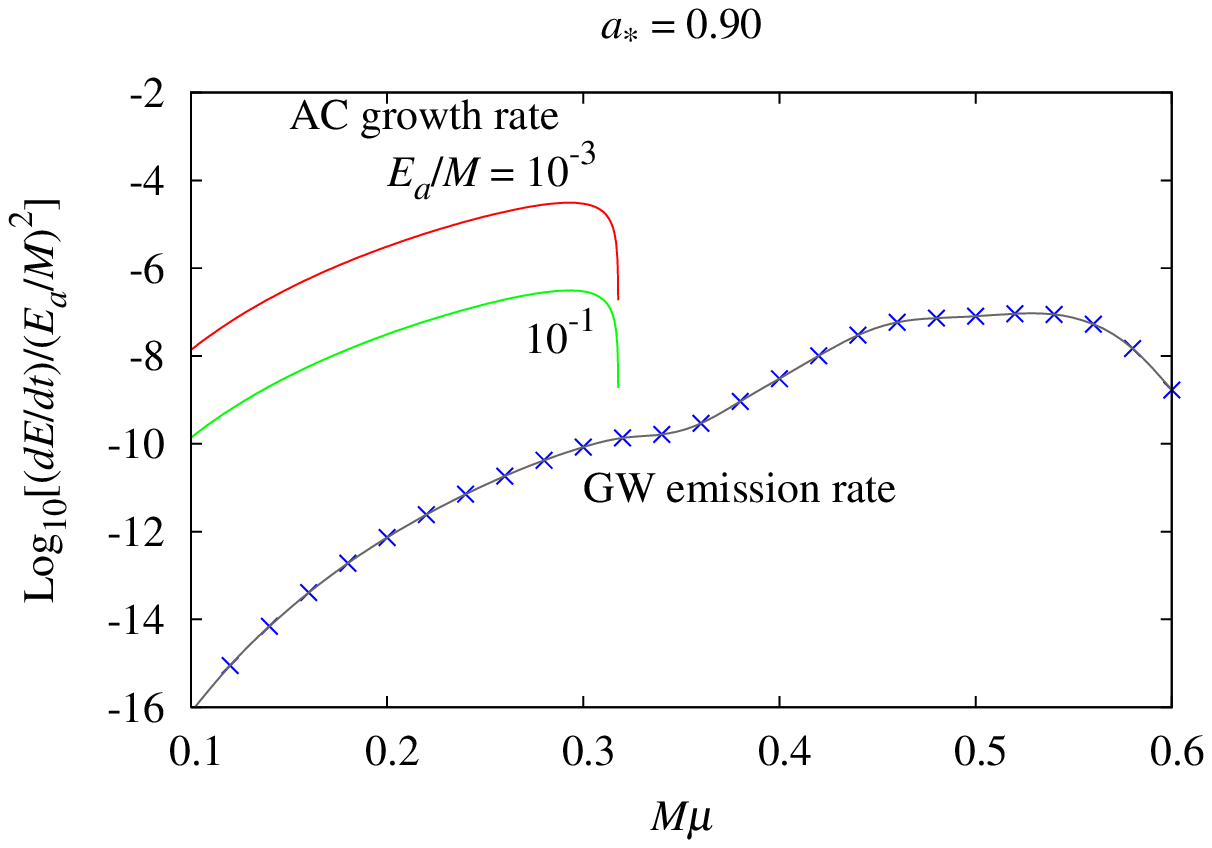}
\includegraphics[width=2.5in]{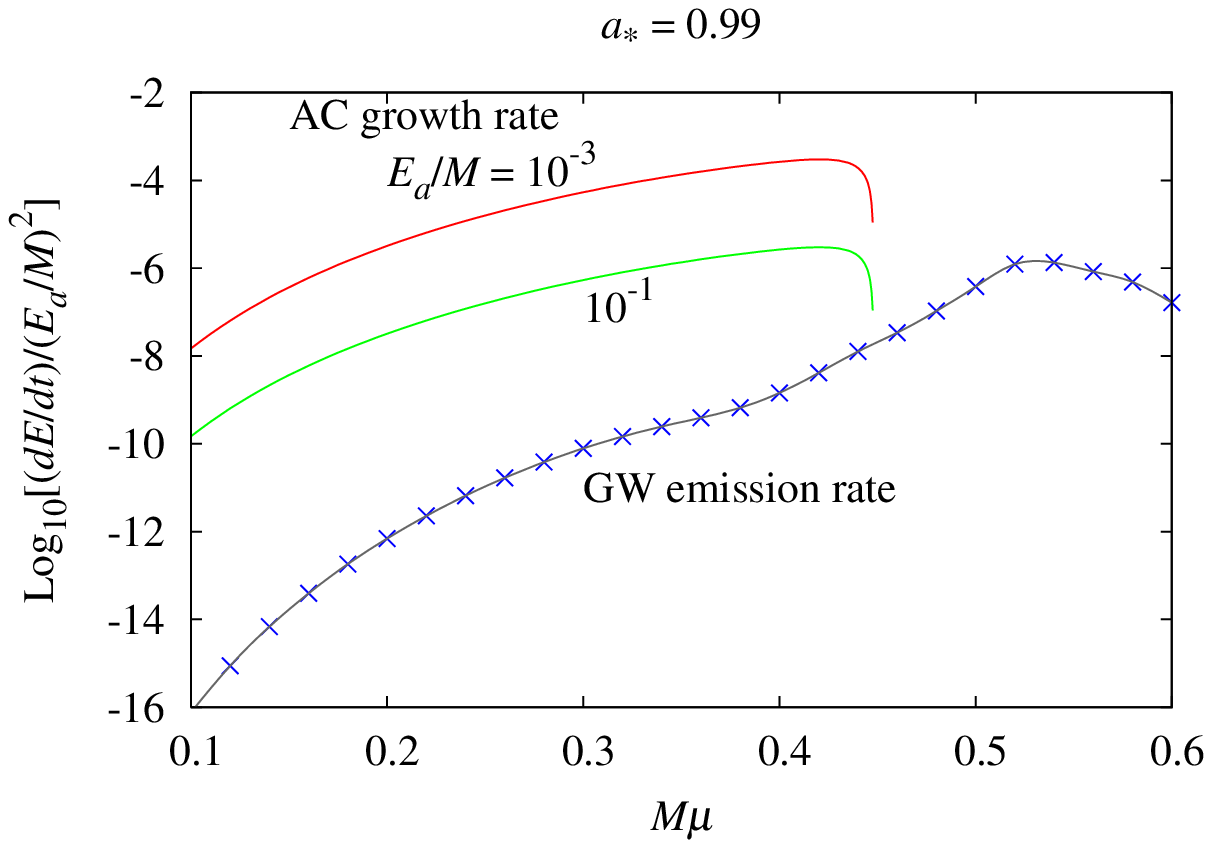}
\caption{Total GW energy emission rate from the axion cloud 
in the $(\ell, m) = (1,1)$ mode and
the energy extraction rate of the axion cloud for the cases of $E_a/M=10^{-3}$
and $10^{-1}$ as functions of $M\mu$. The cases of the BH rotation parameter
$a_*=0.90$ (left) and $0.99$ (right) are shown. The total GW energy emission
rate is evaluated
by summing the first four GW modes $\tilde{\ell} = 2$, $3$, $4$, and $5$.
The energy extraction rate is larger
in all cases.
}
\label{AC_GW_L1M1}
\end{figure}
%

Figure~\ref{AC_GW_L1M1} compares the total GW radiation rate
$dE_{\rm GW}/dt$
and the energy extraction rate $dE_a/dt$
normalized with $(E_a/M)^2$
as functions of $M\mu$ for $a_*=0.90$
and $a_*=0.99$.
Here, the total GW emission rate
is approximated by sum of the GW radiation rates for 
the first four modes with respect to $\tilde{\ell}$,
%
\begin{equation}
\frac{dE_{\rm GW}}{dt}
\approx
\sum_{\tilde{\ell}=\tilde{m}}^{\tilde{m}+3}
\frac{dE_{\rm GW}^{(\tilde{\ell}\tilde{m})}}{dt}.
\end{equation}
%
Since the power dependence on
$(E_a/M)$ of the two rates are different from each other, 
we have to specify this value
to compare them.
As we have done in the flat approximation, we consider the
two cases, $E_a/M=10^{-3}$ and $10^{-1}$.
Each curve of the energy extraction rate
crosses the curve of the GW radiation rate at the point
where the former curve suddenly drops near the threshold
of the superradiant instability. 
Therefore, the GW emission scarcely affects the growth of the
axion cloud due to the superradiant instability.

\subsubsection{$\ell=m=2$}

%
\begin{figure}[tb]
\centering
\includegraphics[width=2.5in]{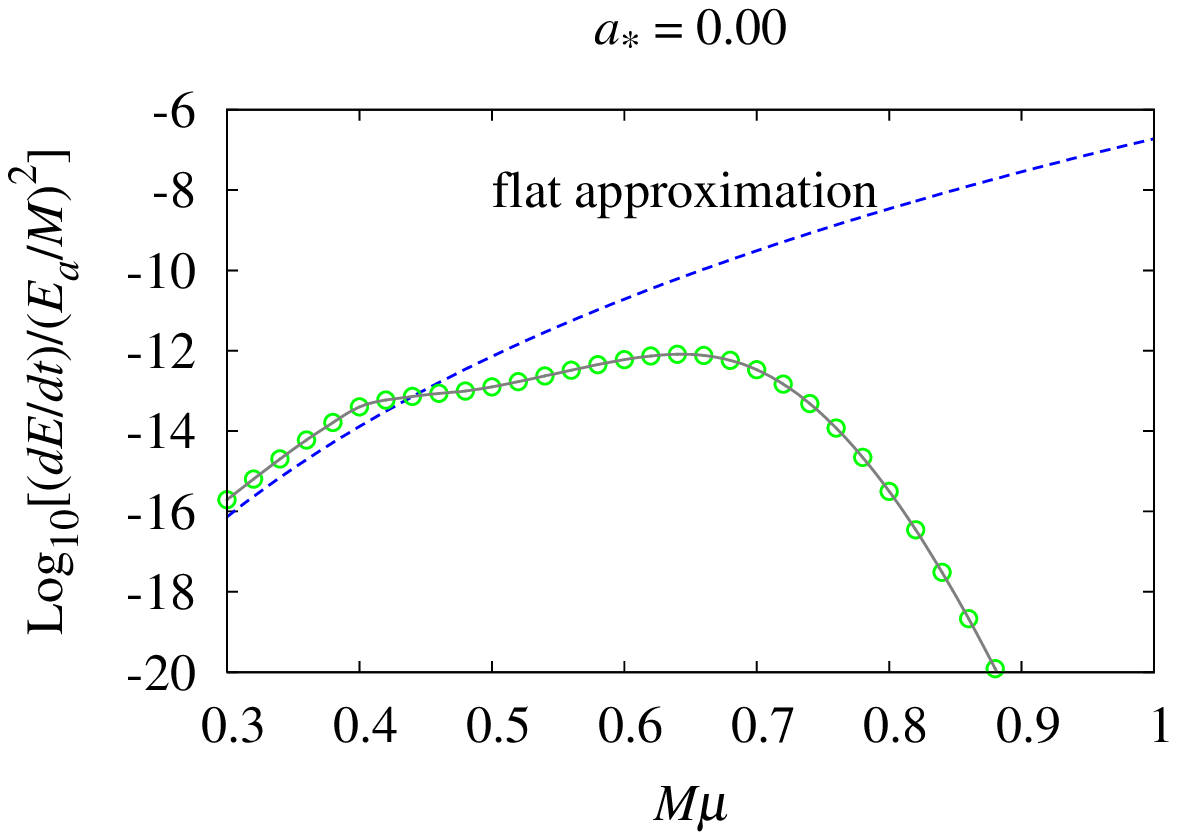}
\includegraphics[width=2.5in]{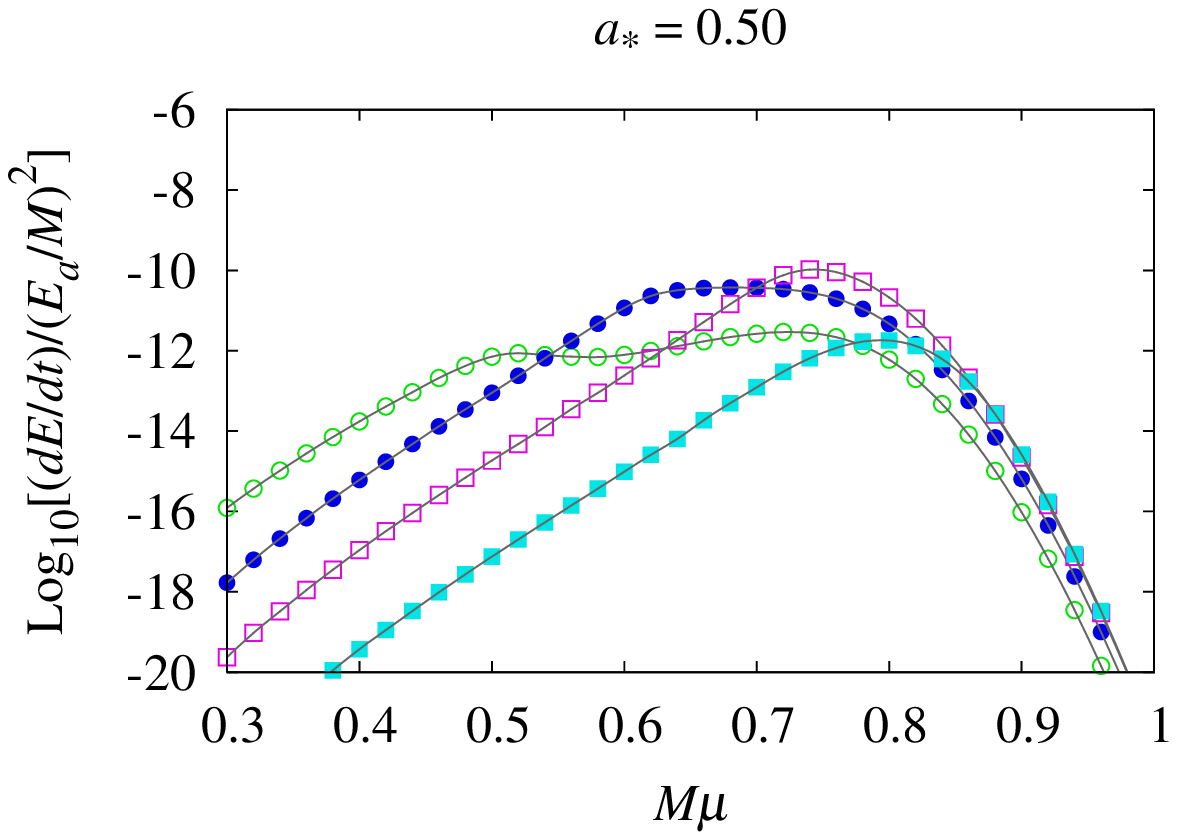}
\includegraphics[width=2.5in]{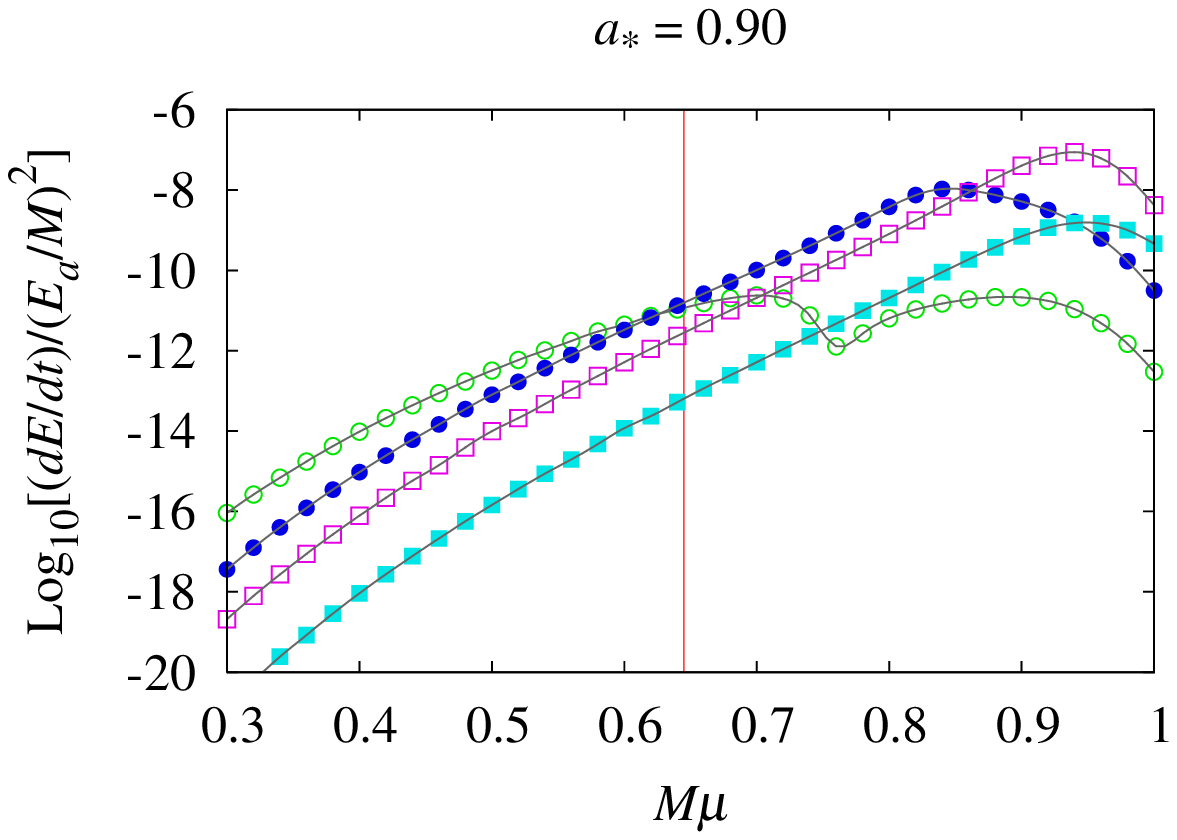}
\includegraphics[width=2.5in]{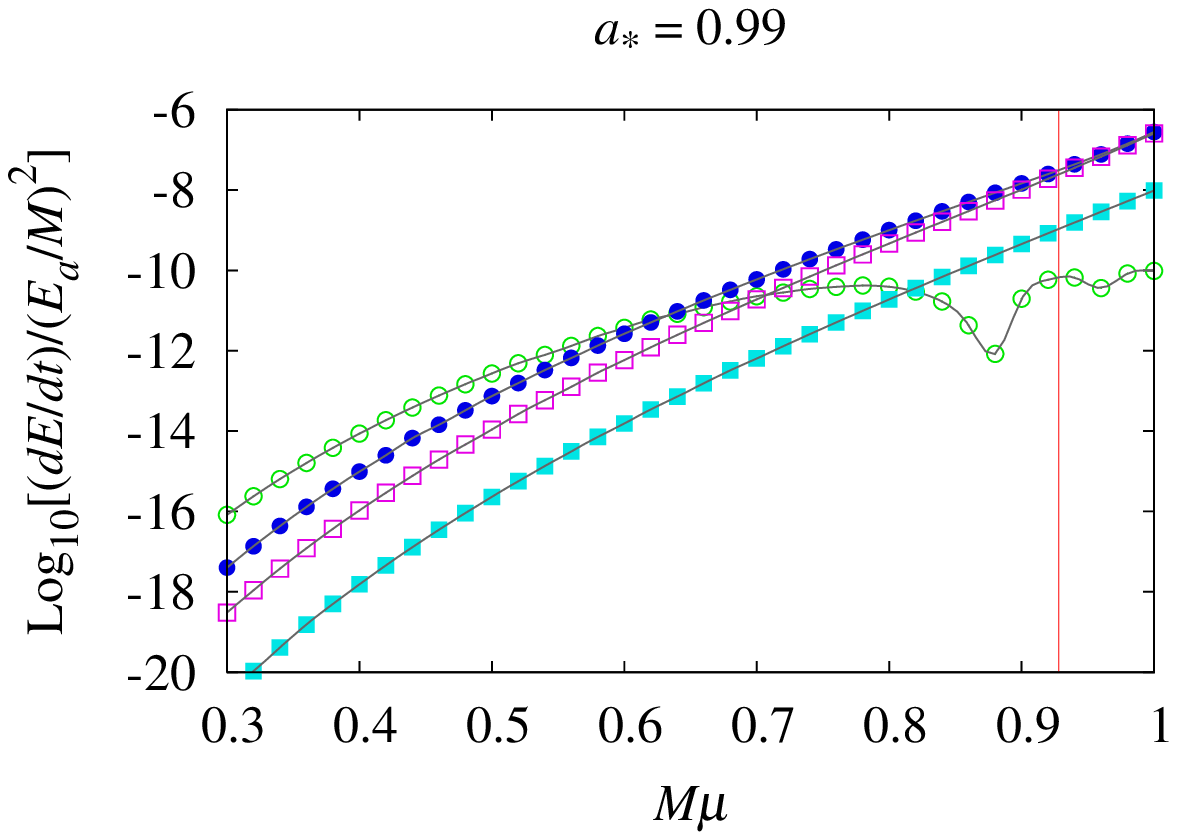}
\caption{The same as Fig.~\ref{GW_rate_mode_L1M1}
for the GW modes $(\tilde{\ell}, \tilde{m}) = (4,4)$ ($\odot$),
$(5,4)$ ($\bullet$),
$(6,4)$ ($\square$),
and $(7,4)$ ($\blacksquare$) from 
the axion cloud in the $\ell = m = 2$ mode.
}
\label{GW_rate_mode_L2M2}
\end{figure}
%

We turn our attention to the results for the axion cloud 
in the $\ell = m = 2$ mode with the radial quantum number $n_r=0$.
Figure~\ref{GW_rate_mode_L2M2} shows the radiation rates
$dE_{\rm GW}^{(\tilde{\ell}\tilde{m})}/dt$
normalized with $(E_a/M)^2$ for the GW modes
$(\tilde{\ell}, \tilde{m})=(4,4)$ (circles, $\odot$),
$(5,4)$ (black circles, $\bullet$),
$(6,4)$ (squares, $\square$), and
$(7,4)$ (black squares, $\blacksquare$)
as functions of $M\mu$ 
for $a_*=0.00$, $0.50$, $0.90$, and $0.99$.
Similarly to the case $\ell = m= 1$,
only the mode $(\tilde{\ell}, \tilde{m})=(4,4)$
is radiated in the case of $a_*=0.00$.
But as the value of $a_*$ is increased,
the contribution of other modes becomes important, and
there are regions where
the modes $(\tilde{\ell}, \tilde{m})=(5,4)$ or $(6,4)$
become dominant.
Again, the effect of nonlinearity with respect to $M\mu$
tends to suppress the GW radiation rate.
Similarly to the case of $\ell=m=1$, the numerical data 
(shown in the panel of $a_*=0.00$)
have the same power dependence $dE_{\rm GW}/dt \propto (M\mu)^{18}$
as the result of the flat approximation 
for $M\mu\ll 1$.

%
\begin{figure}[tb]
\centering
\includegraphics[width=2.5in]{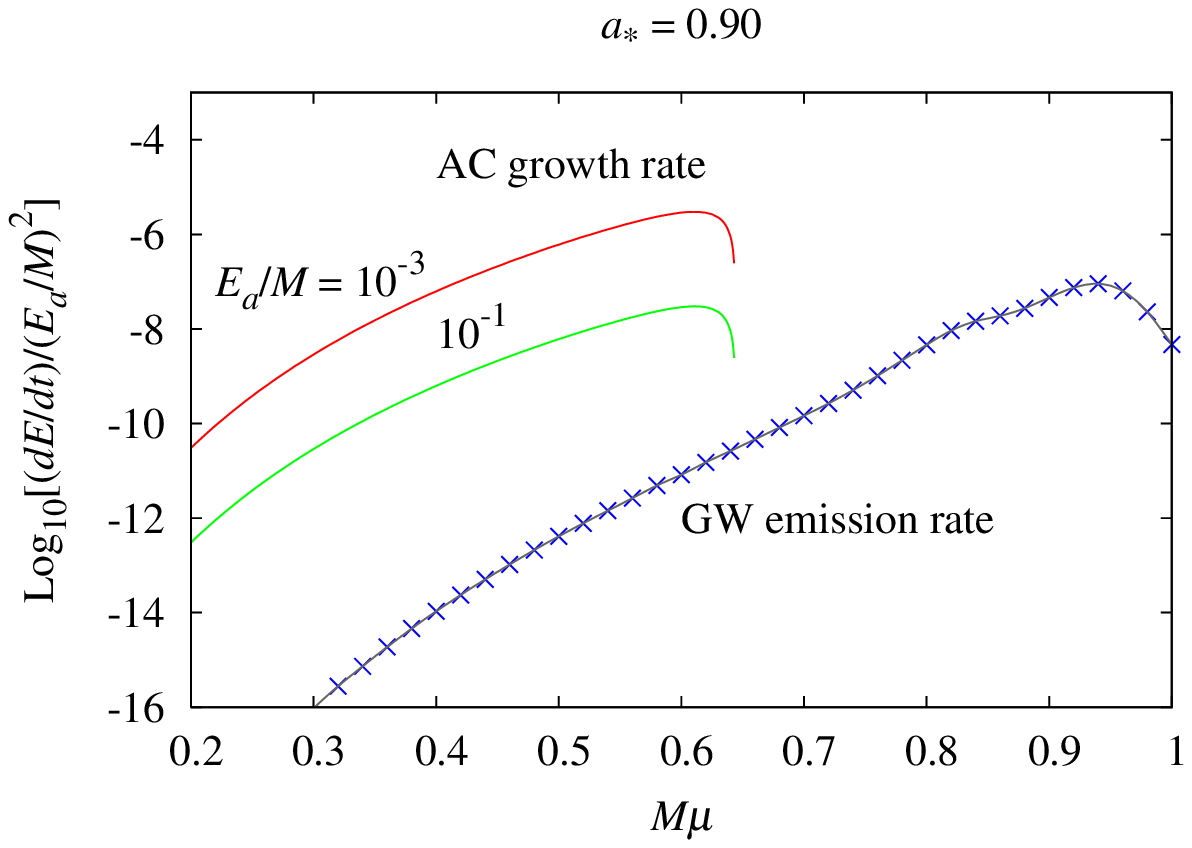}
\includegraphics[width=2.5in]{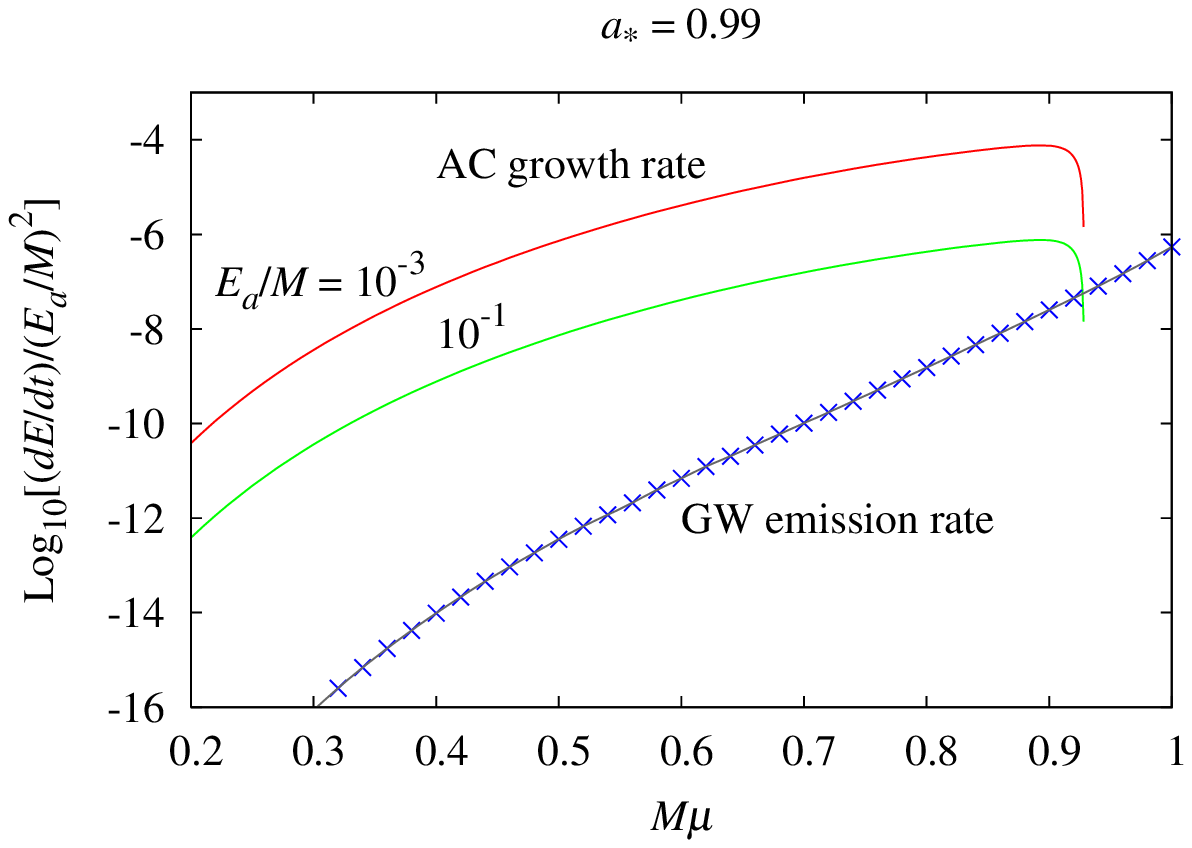}
\caption{The same as Fig.~\ref{AC_GW_L1M1}
but for the axion cloud in the $\ell = m = 2$ mode. 
The total GW energy radiation
rate is evaluated
by summing the first four GW modes $\ell = 4$, $5$, $6$, and $7$.
}
\label{AC_GW_L2M2}
\end{figure}
%

Figure~\ref{AC_GW_L2M2} compares the total GW radiation rate
with the energy extraction rate as functions of $M\mu$ for $a_*=0.90$
and $0.99$.
Here, the total GW radiation rate
is approximated by sum of the GW radiation rates for 
the first four modes with respect to $\tilde{\ell}$,
and the curves of the energy extraction rate for the
cases $E_a/M=10^{-3}$ and $10^{-1}$ are shown. Similarly to
the case $\ell = m = 1$,
the GW radiation rate is much smaller than the energy extraction rate
except for
a very small region near the threshold of the superradiant instability.
Therefore, 
the GW emission scarcely affects the growth of the
axion cloud due to the superradiant instability
also in the case $\ell = m = 2$.

\subsubsection{$\ell=m=3$}

%
\begin{figure}[tb]
\centering
\includegraphics[width=2.5in]{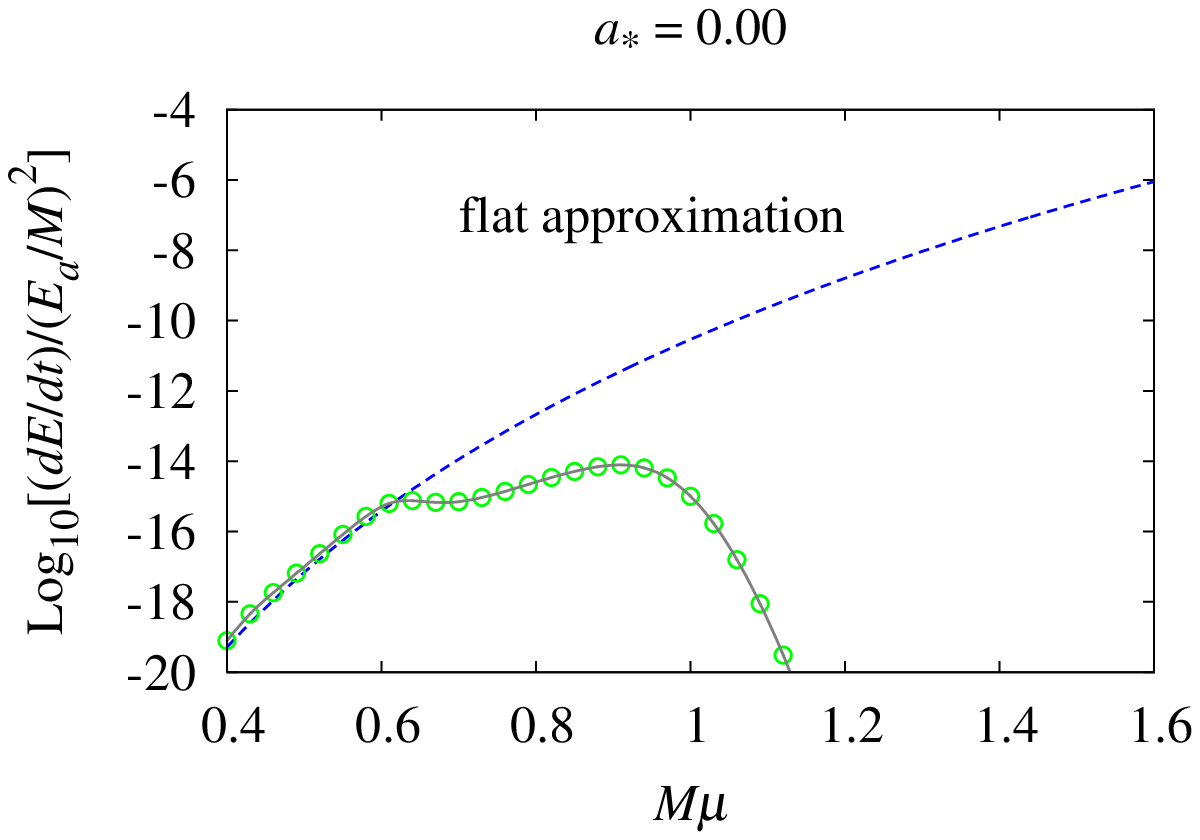}
\includegraphics[width=2.5in]{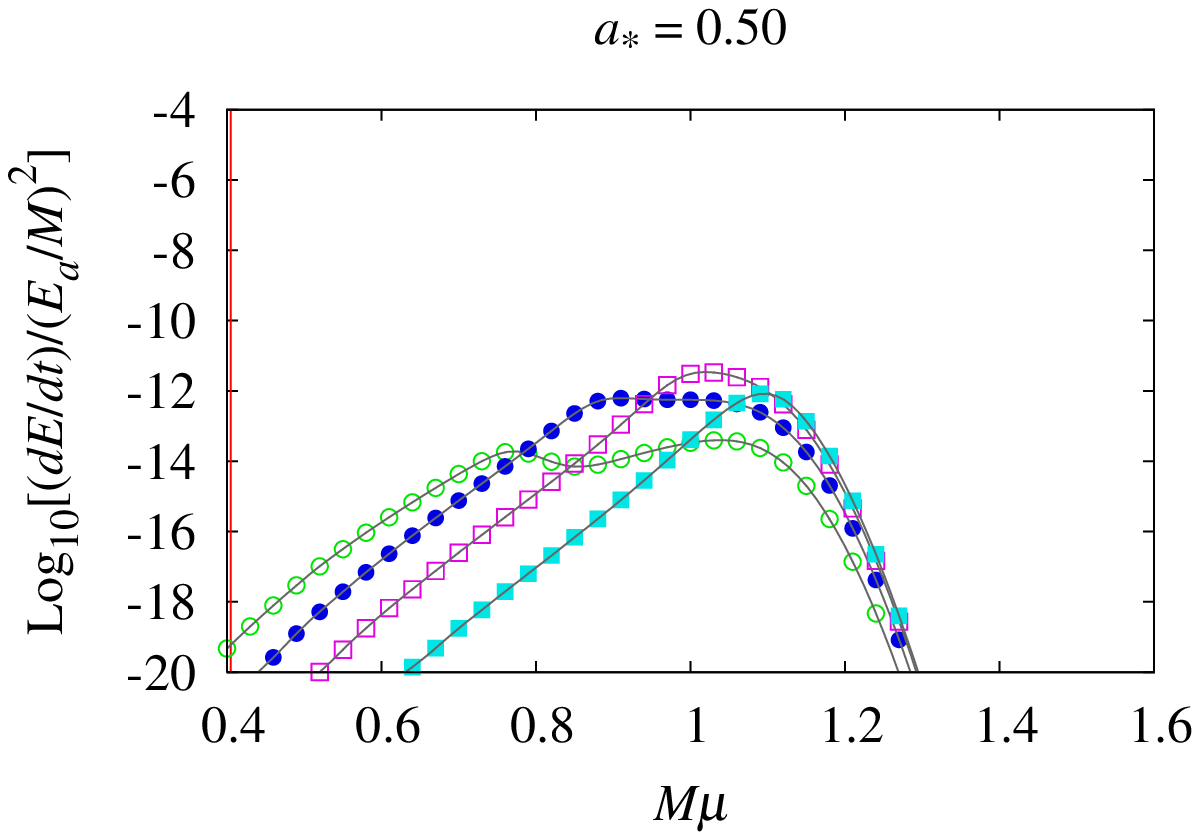}
\includegraphics[width=2.5in]{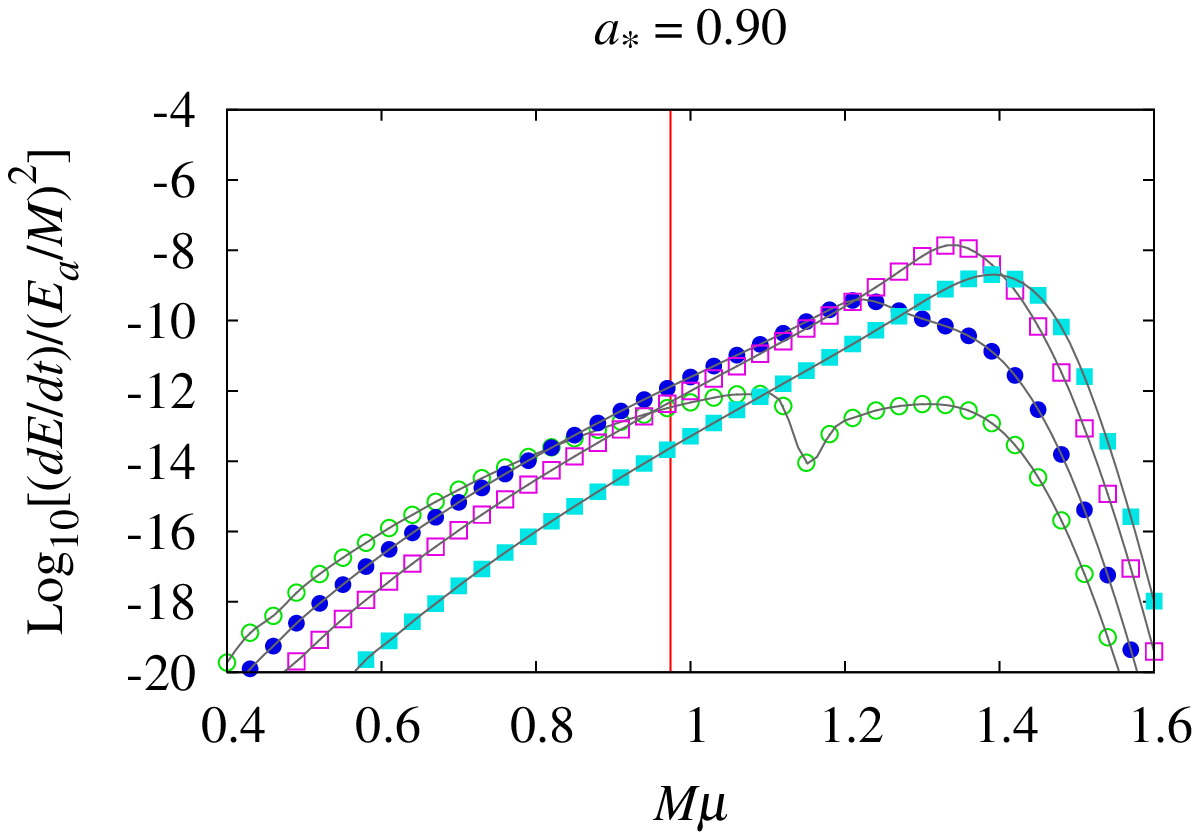}
\includegraphics[width=2.5in]{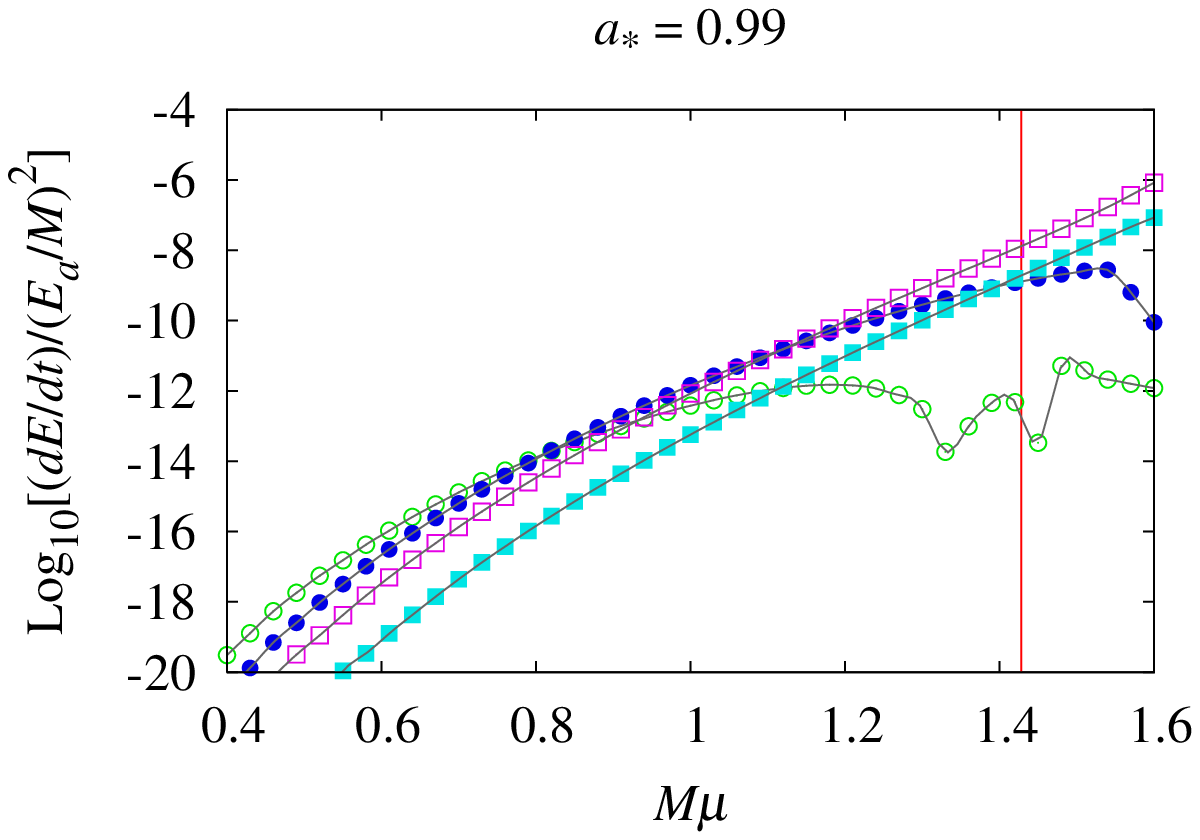}
\caption{The same as Fig.~\ref{GW_rate_mode_L1M1} but 
for the GW modes $(\tilde{\ell}, \tilde{m}) = (6,6)$ ($\odot$),
$(7,6)$ ($\bullet$),
$(8,6)$ ($\square$),
and $(9,6)$ ($\blacksquare$)
from the axion cloud in the $\ell = m = 3$ mode.}
\label{GW_rate_mode_L3M3}
\end{figure}
%

Finally, we show the results for the axion cloud in the $\ell = m = 3$
mode with the radial quantum number $n_r=0$.
Figure~\ref{GW_rate_mode_L3M3} shows the radiation rates
$dE_{\rm GW}^{(\tilde{\ell}\tilde{m})}/dt$
normalized by $(E_a/M)^2$ for the GW modes
$(\tilde{\ell}, \tilde{m})=(6,6)$ (circles, $\odot$),
$(7,6)$ (black circles, $\bullet$),
$(8,6)$ (squares, $\square$), and
$(9,6)$ (black squares, $\blacksquare$)
as functions of $M\mu$
for $a_*=0.00$, $0.50$, $0.90$, and $0.99$.
The general features are same as the previous two cases. 
Although only the mode $(\tilde{\ell}, \tilde{m})=(6,6)$
is radiated in the nonrotating case,
the contribution of other modes become important
as the value of $a_*$ is increased. The GW radiation
is suppressed for a very large value of $M\mu$.
In this case, the numerical data coincide 
well with the flat approximation for $M\mu\ll 1$ as shown in the panel of
$a_*=0.00$.

%
\begin{figure}[tb]
\centering
\includegraphics[width=2.5in]{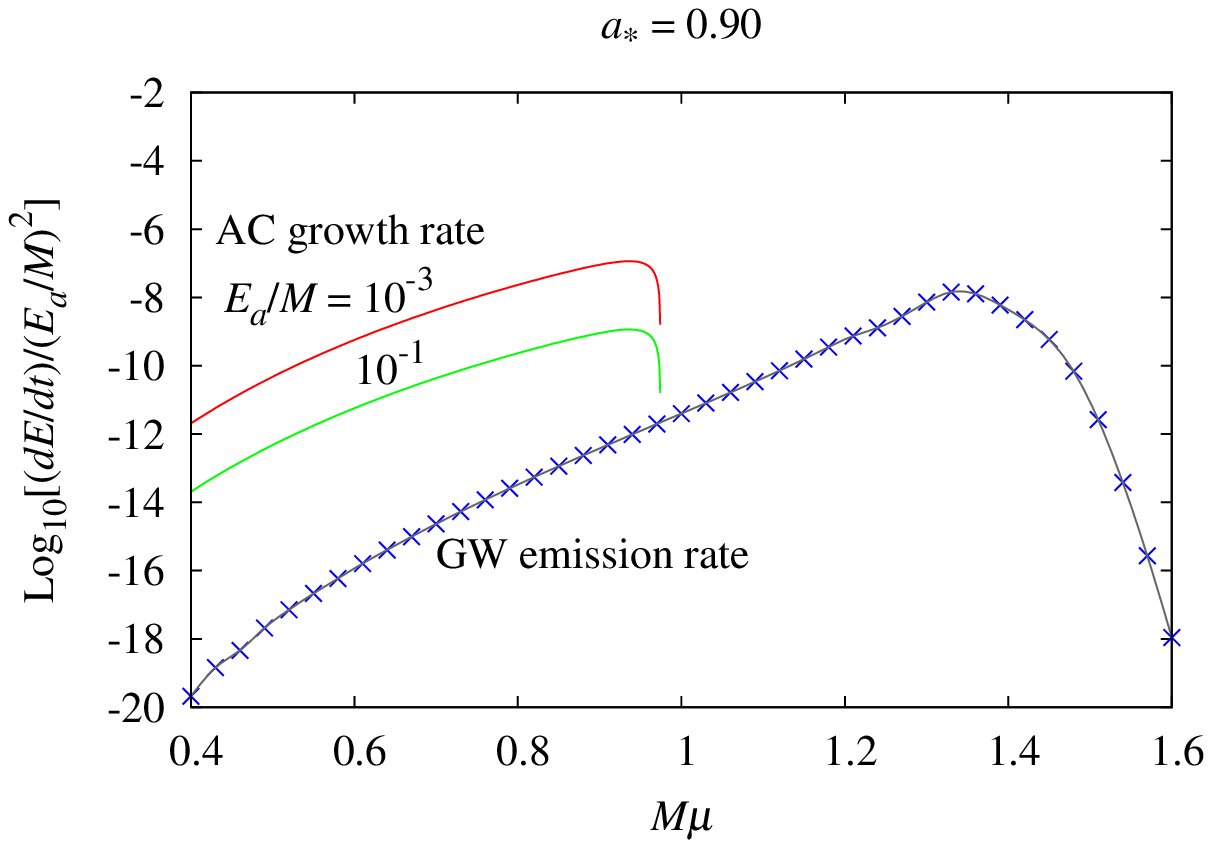}
\includegraphics[width=2.5in]{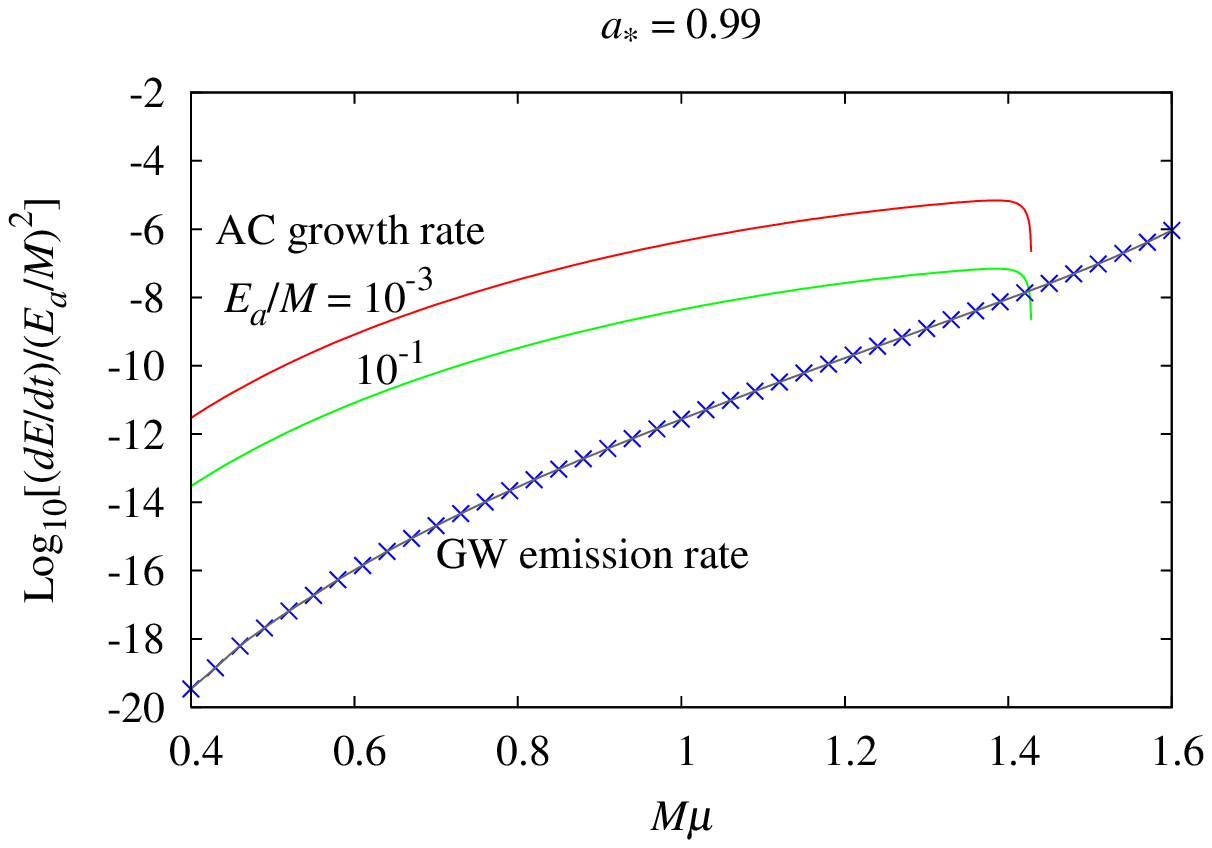}
\caption{The same as Fig.~\ref{AC_GW_L1M1}
but for the axion cloud in the $\ell = m = 3$ mode. 
The total GW energy emission rate is evaluated
by summing the first four GW modes $\ell = 6$, $7$, $8$, and $9$.
}
\label{AC_GW_L3M3}
\end{figure}
%

Figure~\ref{AC_GW_L3M3} compares the total GW radiation rate
(approximated by sum of the radiation rates for 
the first four modes with respect to $\tilde{\ell}$)
and the energy extraction rates
for $E_a/M=10^{-3}$ and $10^{-1}$ as functions of $M\mu$ for $a_*=0.90$
and $0.99$. Again,
the GW radiation rate is much smaller than the energy extraction rate
except for
a very small region near the threshold of the superradiant instability.

\subsection{Summary of the numerical result}

We have calculated the GW radiation rates to infinity,
$dE_{\rm GW}^{\rm (out)}/dt$, as functions of $M\mu$ 
from the axion clouds in the modes
$\ell = m = 1$, $2$, and $3$.
Our numerical data have the same power dependence
as the flat approximation for small values of $M\mu$. 
We compared the numerical results of $dE_{\rm GW}/dt$
with the energy extraction rates 
$dE_a/dt$ of the axion cloud. The value of 
$dE_{\rm GW}^{\rm (out)}/dt$ is much smaller than $dE_a/dt$
for the both cases $E_a/M=10^{-3}$ and $10^{-1}$.
Therefore, similarly to the flat approximation, 
the axion cloud grows by the superradiant 
instability until the effect of the nonlinear self-interaction
becomes important also
when the background spacetime is adopted to
be a rotating Kerr BH spacetime.

%
%
\section{Conclusion}

In this paper, we have studied GW emissions from an axion cloud 
in the superradiant phase around a rotating BH by combining analytic 
methods and numerical calculations. 
First, for $M \mu \ll1$ for which the Newtonian approximation is good, 
we have derived the analytic formula
\eqref{Flat_GW_efficiency_result}--\eqref{CNL} in the flat spacetime 
approximation. This formula can be translated into the expression 
for the observed non-dimensional GW amplitude $h$ as
\begin{equation}
h \sim C^{\rm N} \alpha_g^{2\ell+4} \left(\frac{G M}{d c^2}\right) 
\left(\frac{E_a}{Mc^2}\right);
\quad
C^{\rm N}=\sqrt{2C_{n\ell}}
\end{equation}
with $C_{n\ell}$ given in the Table~\ref{Table:flat-approximation}, 
where $d$ is the distance to the source from us, $\alpha_g$ is defined by
\begin{equation}
\alpha_g = \frac{\mu M}{M_{\rm pl}^2},
\end{equation}
and we have recovered $G$ and $c$.

This formula cannot be used when $\alpha_g$ becomes of the order of unity, 
because the axion cloud approaches the BH horizon and relativistic 
effects become significant. 
Therefore, for the parameter range $\alpha_g \sim 1$, 
we have computed numerically the GW emission rate from the axion cloud 
corresponding to the modes $\ell = m = 1$, $2$, and $3$ 
by solving the perturbation of the Kerr background spacetime exactly. 
As shown in Figs.~\ref{GW_rate_mode_L1M1}, \ref{GW_rate_mode_L2M2}, 
and \ref{GW_rate_mode_L3M3}, the results of the numerical calculations 
indicate that the GW emission rate for each GW mode deviates 
from the above analytic formula when $\alpha_g$ approaches unity, 
and rapidly decreases beyond some critical value of $\alpha_g$ 
due to relativistic effects. This critical value of $\alpha_g$ 
depends on the angular momentum parameter $a_*=J/M^2$ and the 
quantum number $\ell$ characterising the superradiant mode of the 
cloud and increases with $a_*$ and $\ell$. 
Interestingly, up to this critical value, the above analytic 
formula reproduces the numerical results rather well 
when $a_*$ is large if the contributions of all emission modes 
are summed up, although the deviation becomes significant at 
smaller $\alpha_g$ for $a_*\ll 1$.

These GW emission rates were compared with the energy extraction rate 
by the superradiant instability. When the axion decay constant is 
in the range $f_a=10^{16}$--$10^{17}{\rm GeV}$ \cite{Yoshino:2012} 
and the angular momentum is in the range $a_*=0.90$--$0.99$, 
we have found that the GW emission rate is much smaller 
than the energy extraction rate except for a small region 
near the threshold of superradiant instability, 
as illustrated in Figs.~\ref{Flat_GW_AC}, \ref{AC_GW_L1M1}, \ref{AC_GW_L2M2}, 
and \ref{AC_GW_L3M3}.  Therefore, the GW emission does not 
hinder the growth of the axion cloud through the superradiant 
instability for a wide range of the system parameters.

In our previous paper~\cite{Yoshino:2012}, we simulated 
the time evolution of an axion field around a Kerr BH 
taking account of nonlinear self-interaction, and 
showed that ``axion bosenova'' happens
as the result of superradiant instability. 
Our results in this paper indicate that we do not have to change 
this picture even if we take into account the backreaction of 
the GW emission from the axion cloud. Since the GW emission rate 
is sufficiently low, the superradiant instability continues until 
the bosenova happens by nonlinear self-interaction.

Because we can safely declare that bosenova happens, 
our next task is to calculate the GW emission {\it during} bosenova. 
In our previous paper \cite{Yoshino:2012}, we gave 
a preliminary estimation of the emission rate by the quadrupole formula. 
The result can be expressed as
\begin{equation}
h \simeq C^{\rm Q} \alpha_g^{4} \left(\frac{G M}{dc^2}\right) 
\left(\frac{\Delta E_a}{Mc^2}\right);
\quad
C^{\rm Q} =\frac{2\sqrt{10}}{15(\ell+1)^5},
\end{equation}
where $\Delta E_a$ is the change of the axion cloud energy 
during the bosenova collapse. $\Delta E_a$ includes the energy 
that falls into BH, which is around $(0.05$--$0.2) M$. 
Thus, the main difference between the estimation based on the 
flat approximation and that obtained by the quadrupole formula 
comes from the numerical coefficient and the power of $\alpha_g$. 
Here, for small $\ell$, we have found that $C^{\rm N}$ is  of 
a similar order to or less than $C^{\rm Q}$. The power of $\alpha_g$ 
for the flat approximation is always larger than that for 
the quadrupole formula. Hence, the quadrupole formula always gives 
a larger emission rate than that during the superradiant phase, 
which is detectable by the advanced LIGO, advanced VIRGO and KAGRA 
if the source is inside our galaxy. This also implies that bosenova 
can produce strong GW emissions for small $\alpha_g$ if the 
instability growth time is shorter than the age of the central BH.

In the present paper, we have studied the case in which only 
a single mode grows by superradiance. In realistic situations, 
however, there may exist several unstable modes. 
Although the instability growth rates for them are largely different, 
the amplitudes of all of them can become large simultaneously 
if the instability growth times are all shorter than the age of the BH. 
Then, GW emissions in the superradiant phase will be more complicated 
because the emitted GWs will be the superposition of those 
from the two-axion annihilation processes in several bound-state levels 
and from the process of axion level transitions. 
This multi-level occupation will also make the bosenova collapse 
and the associated GW emissions much more complicated. 
Further, we should also carefully estimate the detectability of 
GW emissions in the superradiant phase because it lasts for much 
longer time than the bosenova collapse. 
Clearly, GW emissions in such realistic situations can be correctly 
evaluated only by combining the method developed in the present paper 
and numerical simulations of bosenova in our previous 
paper~\cite{Yoshino:2012}. 
This program is now in progress.

\section*{Acknowledgments}

H.Y. thanks Masato Nozawa, Takahiro Tanaka, Yasumichi Sano, Misao Sasaki,
and Kouji Nakamura for helpful comments.
This work is supported by the Grant-in-Aid for
Scientific Research (A) (22244030).


%

\appendix

%
%
\section{Numerical construction of axion bound states}
\label{Appendix-A}

In this section, we present details of the numerical
construction of the quasibound state of an axion cloud.

\subsection{Equation}

Assuming the form
of Eq.~\eqref{Phi-separation},
$\hat{\Phi} = e^{-i\omega t}R(r)S(\theta)e^{im\phi}$,
the massive Klein-Gordon equation~\eqref{massive-KG}
in the Kerr background spacetime with the metric~\eqref{Metric-Kerr-ST}
is reduced to the following separated two equations:
%
\begin{equation}
\frac{d}{dr}\Delta
\frac{dR}{dr}
+\left[
\frac{K^2}{\Delta}
-\lambda_{\ell m}
-\mu^2r^2
\right]R=0,
\label{Eq:radial_Teukolsky_scalar}
\end{equation}
%
%
\begin{equation}
\frac{1}{\sin\theta}
\frac{d}{d\theta}\sin\theta\frac{dS}{d\theta}
+\left[
-k^2a^2\cos^2\theta-\frac{m^2}{\sin^2\theta}+A_{\ell m}
\right]S=0,
\label{Eq:angular_Teukolsky_scalar}
\end{equation}
%
where definition of $k$ is same as Eq.~\eqref{Def:k},
$k=\sqrt{\mu^2-\omega^2}$,  and
%
\begin{equation}
K=(r^2+a^2)\omega-am,
 \qquad
\lambda_{\ell m}=A_{\ell m}+a^2\omega^2-2am\omega.
\end{equation}
%
The angular equation \eqref{Eq:angular_Teukolsky_scalar}
is identical to the equation for the spin-weighted spheroidal harmonics
\eqref{Eq:angular_Teukolsky_spin} with $s=0$ if we replace
$a^2\tilde{\omega}^2$ with $-k^2a^2$.

\subsection{Angular eigenvalues and angular eigenfunctions}

The eigenvalue $A_{\ell m}$ has to be evaluated approximately
or numerically for $a_*\neq 0$. In this paper, we use the approximate
formula derived by Seidel~\cite{Seidel:1988} (see also \cite{Berti:2005}).
Setting $c^2=-k^2a^2$, Seidel's approximate formula
is given in the series with respect to $c$ up to the sixth order:
%
\begin{equation}
A_{\ell m} \approx A^{(0)}_{\ell m} + A^{(2)}_{\ell m} c^2
+ A^{(4)}_{\ell m}c^4 + A^{(6)}_{\ell m}c^6.
\end{equation}
%
The zeroth-order term is $A^{(0)}_{\ell m}=\ell(\ell+1)$.
The formulas for $A^{(i)}$ have been given for general $\ell$ and $m$
(or more generally for any spins $s$). This gives a fairly good
approximation.

As for the angular function $S=S_{\ell m}(c;\theta)$, we could not
find an approximate formula for general values of $\ell$
and $m$ in existing studies.
For this reason, we derived expansion formulas $S_{\ell m}(c;\theta)$
with respect to $c$ up to the sixth order
for the values of $(\ell, m)$ necessary
for our purpose. For $\ell=m>0$, the expansion formula can be
expressed as
%
\begin{equation}
S_{\ell m} \approx B \sin^\ell \theta
\left(1+a^{(2)}\cos^2\theta +a^{(4)}\cos^4\theta
+a^{(6)}\cos^6\theta \right).
\end{equation}
%
Here, $B$, $a^{(2)}$, $a^{(4)}$, and $a^{(6)}$ are
polynomials of $c$, and they are defined from the angular
equation~\eqref{Eq:angular_Teukolsky_scalar} order by order.

\subsection{Continued fraction method}

Next task is to determine the eigenvalue for $\omega$
that corresponds to the quasibound state.
The asymptotic behavior of $R(r)$ at infinity satisfying
the decaying condition is
%
\begin{equation}
R\sim r^{-1+\frac{M(2\omega^2-\mu^2)}{k}}e^{-kr},
\end{equation}
%
where we assumed $\Re k>0$. On the other hand, the behavior
in the neighborhood of the horizon satisfying the ingoing condition
is given as
%
\begin{equation}
R\sim e^{- i(\omega-m\Omega_H)r_*}
\end{equation}
%
with the tortoise coordinate $r_*$ 
defined in Eq.~\eqref{Def:tortoise_coordinate}
and the angular velocity $\Omega_H = a/(r_+^2+a^2)$ of the Kerr BH.
Dolan \cite{Dolan:2007} assumed that
the radial function is expressed by the following infinite
series:
%
\begin{equation}
R(r) = (r-r_+)^{-i\sigma}(r-r_-)^{i\sigma + \chi-1}
e^{-kr}\sum_{n=0}^{\infty}a_n\left(\frac{r-r_+}{r-r_-}\right)^n,
\label{Radial_infinite_series}
\end{equation}
%
where $r_\pm=M\pm b$ with $b=\sqrt{M^2-a^2}$, and
%
\begin{equation}
\sigma = \frac{2Mr_+}{r_+-r_-}(\omega-m\Omega_H),
\qquad
\chi=\frac{M(2\omega^2-\mu^2)}{k}.
\end{equation}
%
Assuming that the series are convergent, the formula
\eqref{Radial_infinite_series}  satisfies the boundary conditions
above. Substituting the formula \eqref{Radial_infinite_series}
into the radial equation \eqref{Eq:radial_Teukolsky_scalar},
we obtain the three-term recurrence relation,
%
\begin{subequations}
\begin{equation}
a_1=-\frac{\beta_0}{\alpha_0}a_0,
\label{ThreeTerm1}
\end{equation}
\begin{equation}
\alpha_na_{n+1}+\beta_na_n+\gamma_{n}a_{n-1}=0,
\label{ThreeTerm2}
\end{equation}
\end{subequations}
%
where
%
\begin{subequations}
\begin{eqnarray}
{\alpha}_n &=& (n+1)[n+1-2i\sigma],
\\
{\beta}_n &=& -2n^2 + 2n
\left[
-1+2i\sigma - 2bk -\frac{M}{k}(\mu^2-2\omega^2)
\right]+{\beta}_0,
\\
{\gamma}_n &=& (n-1)\left[ n+1 -2i\sigma - \frac{2M}{k}(2\omega^2-\mu^2)
\right]
+{\gamma}_1,
\end{eqnarray}
\end{subequations}
%
with
%
\begin{subequations}
\begin{multline}
{\beta}_0 = a^2k^2 - 2M(M+b)(\mu^2-2\omega^2)
+4M^2\omega^2 - (1-2iM\omega)\left[1+\frac{i}{b}(am-2M^2\omega)\right]
\\
+\left[
-2k-\frac{M}{bk}(\mu^2-2\omega^2)
\right]
\left[
b(1-2iM\omega)+i(am-2M^2\omega)
\right]-A_{\ell m},
\end{multline}
\begin{multline}
{\gamma}_1 = \frac{M^2(\mu^2-2\omega^2)^2}{k^2}
-\frac{iM(\mu^2-2\omega^2)}{bk}\left[-am+2M^2\omega+2ib(1-iM\omega)\right]
\\
+(1-2iM\omega)\left[1+\frac{i}{b}(am-2M^2\omega)\right].
\end{multline}
\end{subequations}
%
Although these formulas are apparently different from the ones presented
in Ref.~\cite{Dolan:2007}, we have checked that they exactly agree.
From the three-term recurrence relations~\eqref{ThreeTerm1}
and \eqref{ThreeTerm2}, we obtain the following relation
of the continued fraction:
%
\begin{equation}
0=\beta_0-\frac{\alpha_0\gamma_1}{\beta_1-}
\frac{\alpha_1\gamma_2}{\beta_2-}
\frac{\alpha_2\gamma_3}{\beta_3-}\cdots.
\end{equation}
%
This gives the equations for $\omega$, and we solved numerically
using the Newton method. Typically, we took account of the
first 1000 terms of the recurrence relation (i.e., we assumed
$a_n=0$ for $n\ge 1000$).
The result is presented in Fig.~\ref{AC_growthrate}.

\subsection{Radial function}

Once the eigenfrequency $\omega$ for the quasibound state is
determined, we can numerically calculate the sequence of
numbers $a_n$ using the recurrence relations~\eqref{ThreeTerm1} and
\eqref{ThreeTerm2}. Then, using the formula~\eqref{Radial_infinite_series},
the radial function $R(r)$ is generated.
Some examples can be found 
in Figs.~\ref{AC_radial_L1M1} and \ref{AC_radial_L2M2}.

%
%
\section{Inner product in the flat approximation}
\label{Appendix-B}

In Sec.~\ref{Sec:III}, we presented the formula for the
gravitational radiation rate. The radiation rate
vanishes for the odd-type modes, and is given by
Eqs.~\eqref{Flat_GW_efficiency_result}--\eqref{CNL}
for the even-type modes. These results 
are derived from Eqs.~\eqref{RadiationEfficiency_Zout}
and \eqref{Zout}
after calculating $\langle u, T\rangle$.
In this section, we sketch the calculation of this inner product
$\langle u, T\rangle$.

For simplicity, we denote the radial function and spin-weighted
spherical harmonics for the Teukolsky function as 
$\tilde{R}:= {}_{+2}R_{\tilde{\ell}\tilde{m}}(r)$ and 
${}_{s}\tilde{Y} := {}_{s}Y_{\tilde{\ell}\tilde{m}}(\theta,\phi)$, respectively. 
On the other hand, $R=R_{\ell m}(r)$ and 
$Y:=Y_{\ell m}(\theta,\phi)$ represent the
radial function and spherical harmonics for the axion field.

Because the background spactime is assumed to be flat,
the Kinnersley tetrad and the Newman-Penrose variables become
\begin{equation}
\ell^\mu = (1,1,0,0), \quad n^\mu = (1/2, -1/2, 0, 0), \quad
m^\mu = (0,0,1,i/\sin\theta)/\sqrt{2}r,
\end{equation}
and
\begin{equation}
\rho = -1/r, \quad \beta=-\alpha=\cot\theta/2\sqrt{2}r,
\quad \mu = -1/2r, \quad
\pi = \tau = \gamma=0.
\end{equation}
Then, the functions appearing in the Chrzanowski 
formula~\eqref{Chrzanowski-formula} are
calculated as
\begin{subequations}
\begin{eqnarray}
\mathcal{A}\left(\tilde{R}~{}_{-2}\tilde{Y}e^{-i\tilde{\omega}t}\right)
&=&
\frac{x^2}{2\tilde{\omega}^2}\tilde{R}\ 
\eth\eth{}_{-2}\tilde{Y}e^{-i\tilde{\omega}t}
\\
\mathcal{A}^*\left(\tilde{R}~{}_{+2}\tilde{Y}e^{-i\tilde{\omega}t}\right)
&=&
\frac{x^2}{2\tilde{\omega}^2}\tilde{R}\ 
\eth^*\eth^*{}_{+2}\tilde{Y}e^{-i\tilde{\omega}t}
\end{eqnarray}
\begin{eqnarray}
\mathcal{B}\left(\tilde{R}~{}_{-2}\tilde{Y}e^{-i\tilde{\omega}t}\right)
&=&
\frac{x^4}{4\tilde{\omega}^2}
\left[
2i\tilde{R}_{,x} 
+\left(-2+\frac{2i}{x}+\frac{\tilde{\ell}^2+\tilde{\ell}-2}{x^2}\right)\tilde{R}
\right]{}_{-2}\tilde{Y}e^{-i\tilde{\omega}t}
\\
\mathcal{B}^*\left(\tilde{R}~{}_{+2}\tilde{Y}e^{-i\tilde{\omega}t}\right)
&=&
\frac{x^4}{4\tilde{\omega}^2}
\left[
2i\tilde{R}_{,x} 
+\left(-2+\frac{2i}{x}+\frac{\tilde{\ell}^2+\tilde{\ell}-2}{x^2}\right)\tilde{R}
\right]{}_{+2}\tilde{Y}e^{-i\tilde{\omega}t}
\end{eqnarray}
\begin{eqnarray}
\mathcal{C}
\left(\tilde{R}~{}_{-2}\tilde{Y}e^{-i\tilde{\omega}t}\right)
&=&
\frac{x^3}{\sqrt{2}\tilde{\omega}^2}
\left(i\tilde{R}+\tilde{R}_{,x}+\frac{2}{x}\tilde{R}\right)\ 
\eth{}_{-2}\tilde{Y}e^{-i\tilde{\omega}t}
\\
\mathcal{C}^*
\left(\tilde{R}~{}_{+2}\tilde{Y}e^{-i\tilde{\omega}t}\right)
&=&
\frac{x^3}{\sqrt{2}\tilde{\omega}^2}
\left(i\tilde{R}+\tilde{R}_{,x}+\frac{2}{x}\tilde{R}\right)\ 
\eth^*{}_{+2}\tilde{Y}e^{-i\tilde{\omega}t}
\end{eqnarray}
\end{subequations}
On the other hand, the necessary components
of the energy-momentum tensor, 
Eqs.~\eqref{EnergyMomentum-nn}--\eqref{EnergyMomentum-nms}, become
\begin{subequations}
\begin{eqnarray}
\hat{T}_{\mu\nu}m^\mu m^\nu &=&
\frac{\tilde{\omega}^2}{4x^2}R^2 \left(\eth Y\right)^2e^{-2i\omega t},\\
\hat{T}_{\mu\nu}m^{*\mu} m^{* \nu} &=&
\frac{\tilde{\omega}^2}{4x^2}R^2 \left(\eth^* Y\right)^2e^{-2i\omega t},
\end{eqnarray}
\begin{equation}
\hat{T}_{\mu\nu}n^\mu n^\nu =
\frac{\tilde{\omega}^2}{8}\left(\frac{i}{2}R+R_{,x}\right)^2 
Y^2e^{-2i\omega t},
\end{equation}
\begin{eqnarray}
\hat{T}_{\mu\nu}n^\mu m^\nu &=&
\frac{\tilde{\omega}^2}{4\sqrt{2}x}\left(\frac{i}{2}R+R_{,x}\right)R 
\left(\eth Y\right)Ye^{-2i\omega t},\\
\hat{T}_{\mu\nu}n^{\mu} m^{* \nu} &=&
\frac{\tilde{\omega}^2}{4\sqrt{2}x}\left(\frac{i}{2}R+R_{,x}\right)R 
\left(\eth^* Y\right)Ye^{-2i\omega t}.
\end{eqnarray}
\end{subequations}
Here, we used the standard definition for the ``eth'' operators, $\eth$ and
$\eth^*$, which act on the spin-weighted spherical harmonics
as
\begin{subequations}
\begin{eqnarray}
\eth \left({}_sY_{\tilde{\ell}\tilde{m}}\right) &:=& 
-\left(\partial_\theta+\frac{i}{\sin\theta}-s\cot\theta\right)
{}_{s}Y_{\tilde{\ell}\tilde{m}},
\\
\eth^* \left({}_sY_{\tilde{\ell}\tilde{m}}\right) &:=& 
-\left(\partial_\theta-\frac{i}{\sin\theta}+s\cot\theta\right)
{}_{s}Y_{\tilde{\ell}\tilde{m}}.
\end{eqnarray} 
\end{subequations}
It is worth noting the following useful formulas in
order to carry out the angular integrations below: 
\begin{subequations}
\begin{eqnarray}
\eth \left({}_sY_{\tilde{\ell}\tilde{m}}\right) &=& 
+\sqrt{(\tilde{\ell}-s)(\tilde{\ell}+s+1)}{}_{s+1}Y_{\tilde{\ell}\tilde{m}},
\\
\eth^* \left({}_sY_{\tilde{\ell}\tilde{m}}\right) &=& 
-\sqrt{(\tilde{\ell}+s)(\tilde{\ell}-s+1)}{}_{s-1}Y_{\tilde{\ell}\tilde{m}}.
\end{eqnarray} 
\end{subequations}

\subsection{Odd-type modes}

First, we consider the odd-type modes, $P=-1$. In this case, 
the integrations with respect to the angular coordinates $(\theta, \phi)$
that appear in the inner product $\langle u,T\rangle$ are
calculated as
\begin{subequations}
\begin{eqnarray}
\int \left[(\eth^*{}_{-2}\tilde{Y})^*(\eth^*Y)Y
-(\eth^*{}_{+2}\tilde{Y})^*(\eth Y)Y\right]d\Omega
&=&0,
\\
\int \left[({}_{-2}\tilde{Y})^*(\eth^*Y)^2
-({}_{+2}\tilde{Y})^*(\eth Y)^2\right]d\Omega
&=&0,
\\
\int \left[(\eth\eth{}_{-2}\tilde{Y})^*Y^2
-(\eth^*\eth^*{}_{+2}\tilde{Y})^*Y^2\right]d\Omega
&=&0.
\end{eqnarray}
\end{subequations}
Since all the angular integrals become zero, 
we have $\langle u, T\rangle = 0$.
Therefore, odd-type GWs are not radiated.

\subsection{Even-type modes}

Next, we consider the even-type modes, $P=+1$.
The integrations with respect to the angular coordinates $(\theta, \phi)$
that appear in the inner product $\langle u,T\rangle$ are
\begin{subequations}
\begin{eqnarray}
\int \left[(\eth^*{}_{-2}\tilde{Y})^*(\eth^*Y)Y
+(\eth^*{}_{+2}\tilde{Y})^*(\eth Y)Y\right]d\Omega
&=&
-I_{\Omega}\sqrt{(\tilde{\ell}-1)\tilde{\ell}(\tilde{\ell}+1)(\tilde{\ell}+2)},
\\
\int \left[({}_{-2}\tilde{Y})^*(\eth^*Y)^2
+({}_{+2}\tilde{Y})^*(\eth Y)^2\right]d\Omega
&=&
{2\ell(\ell+1)}I_{\Omega}
\sqrt{\frac{\tilde{\ell}(\tilde{\ell}+1)}{(\tilde{\ell}-1)(\tilde{\ell}+2)}},
\\
\int \left[(\eth\eth{}_{-2}\tilde{Y})^*Y^2
+(\eth^*\eth^*{}_{+2}\tilde{Y})^*Y^2\right]d\Omega
&=&
2I_{\Omega}
\sqrt{(\tilde{\ell}-1)\tilde{\ell}(\tilde{\ell}+1)(\tilde{\ell}+2)},
\end{eqnarray}
\end{subequations}
where we introduced the definition
%
\begin{equation}
I_\Omega = \int 
\left({}_{0}Y_{\tilde{\ell}\tilde{m}}\right)^*
\left(Y_{\ell m}\right)^2d\Omega.
\end{equation}
%

Hereafter, we assume the axion cloud to be in the  $\ell = m $ mode.
In this setup, the integral $I_\Omega$ becomes nonzero
if and only if $\tilde{\ell} =\tilde{m}= 2\ell $, 
and its analytic expression for these mode parameters is
%
\begin{equation}
I_{\Omega}
= \frac{\Gamma(2\ell+2)\sqrt{\Gamma(4\ell+2)}}
{4^{2\ell+1}\left[\Gamma(\ell+1)\right]^2\Gamma(2\ell+3/2)}.
\label{I_Omega}
\end{equation}
%
After integrating $\hat{u}^*_{ab}\hat{T}^{ab}$ with respect to the
angular variables $(\theta,\phi)$, we have
\begin{multline}
\int \hat{u}^*_{ab}\hat{T}^{ab}d\Omega 
= 
\frac{I_{\Omega}}{16}
\sqrt{(\tilde{\ell}-1)\tilde{\ell}(\tilde{\ell}+1)(\tilde{\ell}+2)}
\\
\times\left\{
-\frac{2\ell(\ell+1)x^2}{(\tilde{\ell}-1)(\tilde{\ell}+2)}
\left[-2i\tilde{R}^*_{,x}
+\left(-2-\frac{2i}{x}+\frac{\tilde{\ell^2}+\tilde{\ell}-2}{x^2}\right)
\tilde{R}^*\right]R^2
\right.
\\
\left.
-2x^2\tilde{R}^*\left(\frac{i}{2}R+R_{,x}\right)^2
+x^2\left(2i\tilde{R}^*-2\tilde{R}^*_{,x}-\frac{4}{x}\tilde{R^*}\right)
\left(\frac{i}{2}R+R_{,x}\right)R
\right\}.
\label{InnerProduct1}
\end{multline}
Now, we multiply the volume element in the radial direction
$r^2 dr = \tilde{\omega}^{-3}x^2 dx$
and perform the integration. Integrating by parts and 
rewriting with the equation for the radial function of the axion field $R$, 
\begin{equation}
R_{,xx}+\frac{2}{x}R_{,x}
+\left[-\frac{\beta^2}{4}+\frac{\ell(\ell+1)}{x^2}
-\frac{n\beta}{x}
\right]R=0,
\end{equation}
where $\beta$ is a small parameter $\beta:=k/\omega \approx \mu M/n$,
we have
\begin{multline}
\langle u,T\rangle
\approx
\frac{I_{\Omega}}{16}\tilde{\omega}^{-3}
\sqrt{(\tilde{\ell}-1)\tilde{\ell}(\tilde{\ell}+1)(\tilde{\ell}+2)}
\\
\times
\int
\tilde{R}^* x^4
\left\{
\left[\frac{2\ell+1}{2(2\ell-1)}-\frac{2(\ell+1)}{2\ell-1}\frac{i}{x}\right]R^2
-\frac{2i}{2\ell-1} RR_{,x}
-\frac{2n\beta}{x}R
\right\}dx.
\label{InnerProduct-RadialIntegral}
\end{multline}  
Here, we neglected $O(\beta^2)$ terms compared to the
leading order, because
we are interested only in the leading order term of $\langle u,T\rangle$
with respect to $\beta$.
The radial function for the axion field can be expressed as
\begin{eqnarray}
R&=&Ae^{-y}\left[(kr)^{\ell}+q(kr)^{\ell+1}+\cdots\right]
\nonumber\\
&=&
Ae^{-(\beta/2)x}\left(\frac{\beta}{2}\right)^{\ell}\left[
x^{\ell}
+\frac{q\beta}{2} x^{\ell+1}+\cdots\right]
\label{R_expand}
\end{eqnarray}
with
\begin{equation}
A = -\frac{\sqrt{2E_a}}{\omega}(2k)^{3/2}
\sqrt{\frac{(n-\ell-1)!}{2n(n+\ell)!}}
\frac{2^\ell(n+\ell)!}{(n-\ell-1)!(2\ell+1)!},
\label{Def:A}
\end{equation}
\begin{equation}
q =\frac{\ell-n+1}{\ell+1}.
\label{Def:q}
\end{equation}
%
Substituting Eq.~\eqref{R_expand} 
into Eq.~\eqref{InnerProduct-RadialIntegral}, we have
\begin{multline}
\langle u,T\rangle
=\frac{I_\Omega}{16}\tilde{\omega}^{-3}
\sqrt{(\tilde{\ell}-1)\tilde{\ell}(\tilde{\ell}+1)(\tilde{\ell}+2)}A^2
\left(\frac{\beta}{2}\right)^{2\ell}
\left\{
\frac{2\ell+1}{2(2\ell-1)}f^{(1)}(\beta)
\right.
\\\left.
-
\frac{2(2\ell+1)i}{2\ell-1}f^{(0)}(\beta)
+
\frac{(2\ell+1)q}{2(2\ell-1)} \beta f^{(2)}(\beta)
+
\frac{[1-(4\ell+3)q]i}{2\ell-1}
\beta
f^{(1)}(\beta)
-
2n\beta
f^{(0)}(\beta)
\right\},
\end{multline}
where we defined the formula
\begin{equation}
f^{(i)}(\beta) := \int_0^{\infty}\tilde{R}^*x^{2\ell+3+i}e^{-\beta x}dx.
\end{equation}
The integration of $f^{(0)}$ can be performed analytically as
\begin{eqnarray}
f^{(0)}
&=& 
\frac{\Gamma(4\ell+2)}{(\beta-i)^{4\ell+2}}
\left[1+\frac{4}{(\beta-i)^2}\right]^{1/2-\ell}
\exp\left[-i(2\ell-1)\arctan\left(\frac{2}{\beta-i}\right)\right]
\nonumber\\
&=& \Gamma(4\ell+2)
\left[1-4i\beta -(2\ell+9)\beta^2+\cdots\right],
\end{eqnarray}
and $f^{(i)}$ with $i>0$ are given by differentiating $f^{(0)}$ as
\begin{equation}
f^{(i)} = (-1)^{i}\frac{d^if^{(0)}}{d\beta^i}.
\end{equation}
Substituting these formulas, we find that
the terms of $O(\beta^{2\ell})$ cancel out. Therefore, the 
order of the inner product becomes $O(\beta^{2\ell+1})$,
and the final result is
\begin{equation}
\langle u,T\rangle
=\frac{I_\Omega}{8}\tilde{\omega}^{-3}
\sqrt{(\tilde{\ell}-1)\tilde{\ell}(\tilde{\ell}+1)(\tilde{\ell}+2)}A^2
\frac{n}{\ell+1}\Gamma(4\ell+2)\left(\frac{\beta}{2}\right)^{2\ell+1}.
\label{InnerProduct2}
\end{equation}
Substituting this formula with Eqs.~\eqref{I_Omega}, \eqref{Def:A}
and \eqref{Zout} into Eq.~\eqref{RadiationEfficiency_Zout}, 
we find the formula for the GW energy emission rate,
Eqs.~\eqref{Flat_GW_efficiency_result}--\eqref{CNL}.


\end{document}